\documentclass[usenatbib]{mnras}

\usepackage{natbib}
\usepackage{delarray}
\usepackage{color}
\usepackage{amsmath}
\usepackage{amssymb}
\usepackage{here}
\usepackage{graphicx}
\usepackage[utf8]{inputenc}
\usepackage{float}
\usepackage{epsfig}
\usepackage{tensor}
\usepackage{xspace}
\usepackage{multirow}

\usepackage{newtxtext,newtxmath}

\usepackage[T1]{fontenc}
\usepackage{ae,aecompl}

\newcommand{\mach}{\mathcal{M}}

\newcommand{\be}{\begin{equation}} \newcommand{\ee}{\end{equation}}

\newcommand{\dderiv}{\mathrm{d}}
\newcommand{\vvector}{\mathbf{v}}
\newcommand{\Bvector}{\mathbf{B}}

\newcommand{\acknowledgments}{\begin{small}\section*{Acknowledgments}\end{small}}
\newcommand\altaffilmark[1]{$^{#1}$}
\newcommand\altaffiltext[1]{$^{#1}$}

\newcommand{\frameworkrad}{Paper I\xspace}
\newcommand{\correlationpaper}{Paper II\xspace}

\newcommand{\myquote}[1]{``#1''}

\usepackage[normalem]{ulem}

\title[Universal Scaling Relations in Scale-Free Structure Formation]{Universal Scaling Relations in Scale-Free Structure Formation}

\author[Guszejnov, Hopkins \&\ Grudi{\'c}]{
\parbox[t]{\textwidth}{ D\'avid Guszejnov\altaffilmark{1}\thanks{E-mail:guszejnov@caltech.edu}, Philip F. Hopkins\altaffilmark{1} and Michael Y. Grudi{\'c} \altaffilmark{1}}
\vspace*{6pt} \\
\altaffiltext{1}{TAPIR, MC 350-17, California Institute of Technology, Pasadena, CA 91125, USA} \\
}

\date{To be submitted to MNRAS, \today \vspace{-0.6cm}}
\begin{document}
\maketitle
\label{firstpage}

\begin{abstract}
A large number of astronomical phenomena exhibit remarkably similar scaling relations. The most well-known of these is the mass distribution $\dderiv N/\dderiv M\propto M^{-2}$ which (to first order) describes stars, protostellar cores, clumps, giant molecular clouds, star clusters and even dark matter halos. In this paper we propose that this ubiquity is not a coincidence and that it is the generic result of scale-free structure formation where the different scales are uncorrelated. We show that all such systems produce a mass function proportional to $M^{-2}$ and a column density distribution with a power law tail of $\dderiv A/\dderiv \ln\Sigma\propto\Sigma^{-1}$. In the case where structure formation is controlled by gravity the two-point correlation becomes $\xi_{2D}\propto R^{-1}$. Furthermore, structures formed by such processes (e.g. young star clusters, DM halos) tend to a $\rho\propto R^{-3}$ density profile. We compare these predictions with observations, analytical fragmentation cascade models, semi-analytical models of gravito-turbulent fragmentation and detailed \myquote{full physics} hydrodynamical simulations. We find that these power-laws are good first order descriptions in all cases.
\end{abstract}

\begin{keywords}
stars: formation -- galaxies: star clusters: general -- turbulence -- galaxies: star formation -- cosmology: theory -- ISM: structure
\vspace{-1.0cm}
\end{keywords}

\section{Introduction}\label{sec:intro}

It is well known that the physics of the interstellar medium and star formation  are very complex, involving turbulence, gravity, radiation and chemistry. Despite this complexity a number of physical quantities show scale-free, power-law-like behavior over a large dynamic range: 
\begin{itemize}
\item The initial mass function (IMF) of stars in different regions of the MW and in extragalactic sources is found be close to a power-law for high mass stars with a slope of approximately -2.35 \citep[e.g.][]{Salpeter_slope,IMF_review,IMF_universality}.
\item Similar to the IMF, the mass function of prestellar cores (CMF) in the MW also resembles a power-law at high masses with slopes close to that of Salpeter (e.g. \citealt{Sadavoy_observed_CMF}). 
\item The mass function of clumps (stellar-mass sized condensations of dust and gas) in molecular clouds exhibits power-law distribution with an inferred slope close to -2 \citep[see][]{Kramer_1998_clumps_MF,Johnstone_Bally_2006_clump_MF}.
\item The mass function of giant molecular clouds (GMC) in the MW is also found to be close to a power-law at high masses with a slope somewhat shallower than the canonical IMF value \citep[e.g.][]{Rosolowsky_2005_GMC}, but there can be significant variation with environment (e.g. see \citealt{Colombo_2014_GMC_survey} where the exponents vary between -1.5 and -2.5).
\item The initial mass function of star clusters exhibits a similar power-law behavior with an inferred slope of -2 \citep[e.g.][]{Zhang_Falll_1999,Bik_2003_cluster_IMF,Fall_2012_similarities}.
\item The dark matter halo mass distribution is expected to be close to $\dderiv N/\dderiv M\propto M^{-2} $\citep{PressSchechter,Warren_2006_DM_MF} over a large dynamic range.
\item The column density PDF of star-forming regions can be roughly approximated with a power law $\dderiv A/\dderiv \ln\Sigma\propto\Sigma^{-\gamma}$ \citep{Kainulainen_power_law_tail,Lombardi_2014_column_density}. At low-to-intermediate densities, this appears to be determined by the global mass profile of the cloud with $\gamma\sim 2-3$ \citep{Schneider_2015}, while in the dense star-forming gas the slope appears to approach $\gamma\sim 1$ \citep{Schneider_two_power_law_2015}. 
\item The stellar two-point correlation function in young star clusters has been measured over a wide dynamic range (about 5 orders of magnitude in radius), with the large-scale behavior of the 2D correlation function similar to a power-law with a slope of -1 \citep[e.g.][]{Simon1997,Hartmann2002,Hennekemper2008, Kraus2008}. Note that it has been shown that very different geometries (e.g., fractal vs spherical) can lead to similar correlation function slopes \citep{Gouliermis_2014}.
\item Similarly the 2D two-point correlation functions of protostellar cores (\citealt{Stanke2006}) has also been measured and found to be consistent with power law slopes of -1 or slightly shallower.
\item Studies have investigated the 2D correlation function of star clusters \citep[see][]{Zhang_Fall_2001_cluster_correlation, grasha_2017_cluster_correlation} in nearby galaxies and found it to be close to a power-law of -1 for young clusters within the scale height of the galactic disk. As star formation predominantly happens in GMCs this implies a similar trend for the correlation function of GMCs.
\item The 2D correlation function of dark matter halos has also been found consistent with a power-law with -1 slope both observationally \citep[e.g.][]{Baugh_1996_DM_correlation,Soltan_2015_galaxy_correlation} and numerically \citep[e.g.][]{Kauffmann_1999_DM_corr_sim}. Numerical studies have shown that on intermediate scales ($\ll 10\,\mathrm{Mpc}<$) these results are independent from the initial density power spectrum.
\item The mass profile of young star clusters exhibits power-law-like behavior, the observed surface density profile at large scales is well approximated by power-laws consistent with a density profile with slopes between -3 and -5 \citep[see][]{Elson1987,Mackey_Gilmore_2003,Mackey_Gilmore_2003_SMC,Portegies_young_clusters}.
\item The density profile of dark matter halos is well described by the NFW profile \citep{NFW_1996} that simplifies to $\rho\propto R^{-3}$ on larger scales.
\end{itemize}

There have been a number of attempts to formulate theories to explain some of these scaling relations. A popular idea for gas clouds is to assume that the formation of these objects is set by the interplay between turbulence and gravity \citep[e.g.][]{Padoan_theory,Padoan_Nordlund_2002_IMF,HC08,HC2009,HC_2013,excursion_set_ism,core_IMF,general_turbulent_fragment,guszejnov_GMC_IMF}. These \emph{gravito-turbulent models} have successfully reproduced the mass functions and even the two-point correlation function above \citep[see][]{Hopkins_clustering,guszejnov_correlation}. These have a number of attractive properties including the natural appearance of the linewidth-size relation \citep{Kritsuk_larson_supersonic_origin}. Another interesting aspect of this approach is the apparent universality one obtains in the supersonic limit where the process becomes an almost self-similar fragmentation cascade, washing out most of the differences between individual models \citep{SF_big_problems}. 

Another popular approach to explain these relations in star formation is to rely on self-similar growth as small \myquote{seeds} grow by accreting from the same mass reservoir (originally proposed by \citealt{Larson_1982} then worked out by \citealt{Zinnecker_1982_competitive_accretion}, see review of \citealt{bonnell_2007_competitive_accretion_imf} and references therein for more details). These \emph{competitive accretion models} rely of gravity and hydrodynamics to show that the features of the initial \myquote{seed} distribution are washed out by accretion leading to a power-law distribution consistent with the IMF. 

Finally, in a somewhat different approach, one can notice that the apparent similarity in the slopes of the mass functions could be explained by a fractal-like, self-similar ISM out of which structures like stars, cores and GMCs form \citep[e.g.][]{Elmegreen_1996_fractal,Elmegreen_1997_IMF_fractal_model, Stutzki_1998_fractal_ISM, Chappell_Scalo_2001_multifractal_ISM}. An important property of these models is that they tie structures of different sizes together (stars, cores, clumps) as their mass distribution is the result of the same fractal ISM structure \citep[e.g.][]{Elmegreen_2002_cloud_MF_fractal}. The density structure predicted by these \emph{fractal ISM models} is in agreement with simulations of supersonic turbulence \citep[e.g.][]{Kristuk_2006_turbulence}. In general these inherently imply an underlying self-similar process, which serves as the main motivation for this paper.

While the models above tried to explain the ISM-related phenomena, there has been a similarly large effort related to the scaling laws of dark matter. The DM halo mass function was first predicted by the random field approach of \cite{PressSchechter} and \cite{Bond_extended_PS}, which is actually the same formalism a number of gravito-turbulent theories for star-formation and the ISM are based on \citep[e.g.][]{excursion_set_ism}. A key feature of these Press-Schechter models is that the phases of the different-scale modes in the density field are uncorrelated, in other words: the different scales are independent (this is the reason one can describe the process as a random walk in Fourier-space).  

Note that these classes of models concentrate on quite different physics but still produce similar scaling relations for the mass functions, density PDFs, correlation functions etc. In this paper we aim to demonstrate that these scalings can be explained to first order by \emph{any} scale-free structure building process with a large dynamic range where the different scales are uncorrelated. We argue this point in Sec. \ref{sec:universal_behaviour} and then show that all the processes listed at the beginning of this section can be described with the same generic hydrodynamical problem. To demonstrate the properties of this problem we concentrate on one of its subclasses: the scale-free fragmentation cascade. We formulate a general description of a fragmentation cascade in Sec. \ref{sec_method} then use it to derive the mass distribution of stars/objects (Sec. \ref{sec:IMF}), their correlation function (Sec. \ref{sec:corr}), the gas density distribution function (Sec. \ref{sec:density}) and the power law tail of the young star cluster mass profile (\ref{sec:cluster}). Then we compare our predictions from the fragmentation cascade model with observed data, the outputs of MISFIT, our semi-analytical simulation of cloud fragmentation (\citealt{guszejnov_GMC_IMF}), and with the results of the detailed multi-physics MHD simulations of \cite{Grudic_2016}.

\section{Cause of Universal Behaviour}\label{sec:universal_behaviour}

In Sec. \ref{sec:intro} we listed a large number of astrophysical objects (stars, molecular clouds, star clusters, DM) that at first glance seem to obey very different physics. Let us first investigate the structures that form out of molecular gas (e.g. stars, cores, GMCs). Since the gas can be described as a fluid it must obey the nonrealtivistic MHD+gravity momentum conservation equation
\begin{eqnarray}
\label{eq:mom_eq}
\frac{\partial }{\partial t}\left(\rho \vvector\right)+\nabla\cdot\left(\rho\vvector\otimes\vvector\right)=\nonumber \\
-\nabla P+\eta\rho\nabla^2\vvector+\left(\zeta+\eta/3\right)\nabla\left(\nabla\cdot\vvector\right)+\frac{1}{\mu_0}\left(\nabla\times\Bvector\right)\times\Bvector-\rho\nabla\Phi,
\end{eqnarray}
where $\rho$, $\vvector$ and $\Bvector$ are the usual density, velocity and magnetic fields while $P$ is the thermal pressure, $\eta$ is the dynamic viscosity, $\zeta$ is the bulk viscosity and $\Phi$ is the gravitational potential. By dividing with the characteristic scales of the system (size: $L_0$, velocity: $v_0$, density: $\rho_0$, sound speed: $c_{s,0}$, Alfv\'en velocity: $v_A$) we can make Eq. \ref{eq:mom_eq} dimensionless:
\begin{eqnarray}
\label{eq:mom_eq_dimless}
\frac{\partial }{\partial \tilde{t}}\left(\tilde{\rho} \tilde{\vvector}\right)+\tilde{\nabla}\cdot\left(\tilde{\rho}\tilde{\vvector}\otimes\tilde{\vvector}\right)=\nonumber\\
-\mach^{-2}\tilde{\nabla} \tilde{P}+\mathrm{Re}^{-1}\tilde{\rho}\nabla^2\tilde{\vvector}+\tilde{\zeta}\tilde{\nabla}\left(\tilde{\nabla}\cdot\tilde{\vvector}\right)+\mach_A^{-2}\left(\tilde{\nabla}\times\tilde{\Bvector}\right)\times\tilde{\Bvector}-\alpha\tilde{\rho}\tilde{\nabla}\tilde{\Phi},
\end{eqnarray}
where $\tilde{t}\equiv t v_0/L_0$, $\tilde{\nabla}\equiv L_0\nabla$, $\tilde{P}\equiv\frac{P}{\rho_0 c_s^2}$, $\tilde{B}\equiv\frac{B}{v_A \sqrt{\rho_0 \mu_0}}$ and $\tilde{\Phi}\equiv\frac{\Phi}{G \rho_0 L_0^2}$ where $G$ is the gravitational constant, while $\mach=v_0/c_{s,0}$ is the Mach number, $\mathrm{Re}\equiv\frac{\rho_0 v_0 L_0}{\nu}$ is the Reynolds number,$\tilde{\zeta}\equiv\frac{\zeta+\eta/3}{\rho_0 v_0 L_0}$, $\mach_A\equiv v_0/v_A$ is the Alfv\'en Mach number and $\alpha\equiv v_0^2/(G \rho_0 L_0^2)$ is the virial parameter.

Let us now look at the process of star and cloud formation in more detail! In a wide dynamic range of this process the flows are supersonic ($\mach\gg 1$) making the pressure term negligible. Magnetic fields are similarly not important on these scales ($\mach_A\gg 1$), as shown both in simulations \citep[e.g.][]{Federrath_sim_2012,Grudic_2016} and (to some extent, on large scales) in observations \citep[see review of ][]{Crutcher_2009_MC_magnetic_fields}. Meanwhile, viscous forces only matter close to the dissipation scale (i.e. $\mathrm{Re}\gg 1$ and $\tilde{\zeta}\ll 1$). This leaves us with the following dimensionless equation:
\be
\label{eq:mom_eq_dimless_final}
\frac{\partial }{\partial \tilde{t}}\left(\tilde{\rho} \tilde{\vvector}\right)+\tilde{\nabla}\cdot\left(\tilde{\rho}\tilde{\vvector}\otimes\tilde{\vvector}\right)=-\alpha^{-1}\tilde{\nabla}\tilde{\Phi}.
\ee
Eq. \ref{eq:mom_eq_dimless_final} describes the motion of a pressureless fluid in a gravitational potential\footnote{Note that supersonic flows are not perfectly pressureless because they create shocks where pressure inevitably becomes important. Nevertheless, the thickness of a shock transition will generally be much smaller than the scales of the flows generating the shocks (represented by the Mach number $\mathcal{M}$) by a factor $\sim \mathcal{M}^{-2}$.}. Note that this equation is completely scale-free (all quantities are normalized) and its only parameter is the dimensionless $\alpha$, the virial parameter that describes the ratio of kinetic to gravitational energy in the system. 

We can do the same exercise for the case of star clusters and dark matter halos. Both can be described by the collisionless Boltzmann equation with self-gravity which we can bring to the dimensionless form  
\be
\label{eq:boltzmann_dimless}
\frac{\partial \tilde{f}}{\partial \tilde{t} }+\tilde{\vvector}\cdot \tilde{\nabla}\tilde{f}-\alpha^{-1} \tilde{\nabla}\tilde{\Phi}\cdot \frac{\partial\tilde{f}}{\partial\tilde{\vvector}}=0,
\ee
where $f$ is the 6D phase space density function. Once again we find that the governing equation of structure formation is scale-free and only depends on the virial parameter $\alpha$ . Note that the fluid equation of Eq.\ref{eq:mom_eq_dimless_final} is just the first moment of the dimensionless Boltzmann equation (Eq. \ref{eq:boltzmann_dimless}).

Now let us concentrate on what kind of processes we are trying to describe. We are interested in how certain astronomical objects form (e.g. DM halos, GMCs, star clusters). These are gravitationally bound objects, which, by definition, means that they have a specific virial parameter (of order unity). Thus, regardless of their inherent differences, the formation of molecular clouds, star clusters and dark matter halos all follow Eq. \ref{eq:boltzmann_dimless} with a similar virial parameter $\alpha$. Because of this attractor these different phenomena produce similar scaling relations as shown later in Sec. \ref{sec:results}.

\subsection{The Importance of Uncorrelated Scales}\label{sec:attractor}

It is clear that the equation that describes the evolution of structure formation (Eq. \ref{eq:boltzmann_dimless} or Eq. \ref{eq:mom_eq_dimless_final}) is scale-free, in other words: it does not know about the absolute size of the system. But it is also important to look at whether the process has \myquote{memory}, i.e. does a structure remember its progenitor?

By looking at Eq. \ref{eq:boltzmann_dimless} we find that it has actually two time scales: the \emph{crossing time} scale $t_{\rm cross}(L)\sim L/v_0=\frac{L}{L_0}t_0$ and the \emph{gravitational/freefall time} scale $t_{ff}\sim\left(G\rho_0\right)^{-1/2}\sim \alpha^{1/2} t_0\approx t_0$. For marginally self-gravitating structures ($\alpha\sim 1$) the crossing time is shorter than the freefall time on all scales except the largest where they are equal. This means that during the evolution of a self-gravitating object there is more than enough time for mixing on small scales. Since Eq. \ref{eq:boltzmann_dimless} is highly nonlinear (e.g. admits turbulence) this mixing effectively erases the details of the initial conditions ($\tilde{\rho}$, $\tilde{\vvector}$ etc.) on smaller scales. Thus, as we argue below, the initial conditions for a newly formed self-gravitating substructure (whose evolution is also described by Eq. \ref{eq:boltzmann_dimless}) should be independent (at least to leading order) from the initial conditions of its progenitor. 

Another way to say this is that, if we consider some sub-volume $\Omega$ of the parent system which is somehow isolated from its parent (by, say, collapsing under self-gravity), the initial micro-state (exact spatially-dependent values of $\tilde{\rho}({\bf x},\,t=t_{0})$, $\tilde{\bf v}({\bf x},\,t=t_{0})$, etc.) will be \myquote{wiped out} by small-scale (e.g.\, turbulent) motions, on a timescale which is small compared to the global evolution timescale (the dynamical time) of $\Omega$. The {\em statistical} distribution of properties can only depend on the one governing parameter of the equations, $\alpha$ -- so sub-systems with the same $\alpha$ must be statistically identical (after this initial short time), up to the overall normalization/units of the system (e.g.\, its size). In other words: if the different scales are uncorrelated, the statistics of objects of different generations are the same\footnote{Note that the cosmological models for dark matter halos \citep{PressSchechter} and the excursion set models of turbulent fragmentation \citep{excursion_set_ism} all rely on the assumption of uncorrelated scales.}.

We could, conceivably, imagine a process which ``selects" a different value of $\alpha$ for each ``level'' in scale (say, each time one moves in scale, $\alpha$ doubles). This would imprint a systematic difference in the statistics of small-scale systems as compared to large-scale systems. However, the physics of interest for the properties we study here is gravity, which (by definition) selects the {\em same} $\alpha \sim 1$ at all scales -- if we define ``structures'' by self-gravitating or collapsing objects, or fragments, or merging agglomerations, then they must be at similar $\alpha$. Given the assumptions above, this means that each sub-structure must, in turn, have similar statistical properties to its parent. 

Consider the specific example of fragmentation where a large structure repeatedly breaks up into smaller objects (or the opposite where small objects join to form larger structures, i.e. hierarchical merging), but leaves some mass "behind" at each scale. Since the process is scale-free the amount of mass \myquote{left} at each scale has to be some fraction of the current mass, but because the process has no memory it must be the same fraction at every mass scale. If the process has a wide dynamic range than it follows that it leaves only a small fraction of its mass at every scale so the absolute amount of mass is roughly equal at the different scales. This leads to 
\be 
\label{eq:main_scaling}
\frac{\dderiv \log M_{\rm total}}{\dderiv \log x}=\epsilon\sim \rm{const.}\ll 1,
\ee
where $x$ is some physical quantity in which the process is moving up/down in scale (e.g. size, mass). Due to the scale-free nature of the problem all physical quantities are power-laws of each other thus Eq. \ref{eq:main_scaling} leads to a large number of scaling laws. Let us further simplify the expression in the limit $\epsilon \rightarrow 0$, obtaining the following scaling laws:
\begin{eqnarray}
\mathrm{Mass\,Function:\,}\frac{\dderiv \log M_{\rm total}}{\dderiv \ln M}= 0\rightarrow \frac{\dderiv N}{\dderiv M} \propto M^{-2}\\
\mathrm{Density\,PDF:\,}\frac{\dderiv \log M_{\rm total}}{\dderiv \ln \rho}= 0\rightarrow \frac{\dderiv V}{\dderiv \ln \rho} \propto \rho^{-1}\\
\mathrm{Column\,Density\,PDF:\,}\frac{\dderiv \log M_{\rm total}}{\dderiv \ln \Sigma}=  0\rightarrow \frac{\dderiv A}{\dderiv \ln \Sigma} \propto \Sigma^{-1}\label{eq:last_scaling}
\end{eqnarray}
A more rigorous derivation of the above scalings is presented in Sec. \ref{sec:results} for the special case of a fragmentation cascade.

Note that the above scaling relations have been derived numerous times for different systems using very different methods (e.g. using random fields for the DM halo mass function \citealt{PressSchechter} or competitive accretion for the IMF \citealt{bonnell_2007_competitive_accretion_imf}). While these models seem to describe very different physics, they can be all labeled as scale-free structure formation with uncorrelated scales thus they will tend towards the scaling relations of Eqs. \ref{eq:main_scaling}-\ref{eq:last_scaling}.

\section{General Model for Scale-Free Fragmentation}\label{sec_method}

In this section we develop a simple but general model for self-similar fragmentation cascades which describe a significant portion of the physical phenomena we list in Sec. \ref{sec:intro} (e.g. formation of stars, cores, clumps). Our aim is to clearly demonstrate for this subclass that the scaling relations of Eq. \ref{eq:main_scaling}-\ref{eq:last_scaling} are inherent in these processes. In the model we present here we build on the models presented in \citealt{guszejnov_feedback_necessity} and \citealt{guszejnov_correlation} henceforth referred to as \frameworkrad and \correlationpaper.

\begin{figure}
\begin {center}
\includegraphics[width=\linewidth]{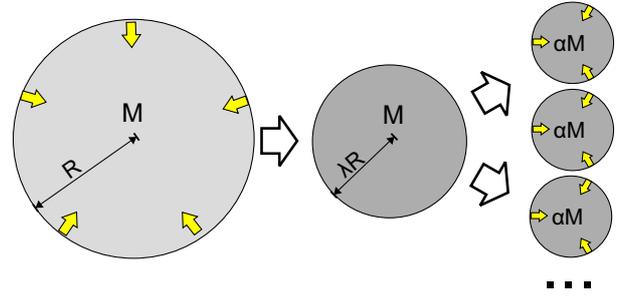}
\caption{Cartoon illustrating the representative toy model for fragmentation, which we use in the text to demonstrate how scale-free fragmentation processes produce universal scalings. In the toy model, every cloud contracts by a factor $\lambda$ (i.e. from initial radius $R$ to $\lambda\,R$), before breaking into some number of sub-fragments, each with mass fraction $\kappa$ of the parent clump mass. The fragments have the same density as their parent. A small fraction $\epsilon \ll 1$ of the clouds are \myquote{left behind} (or collapse without fragmentation) at each level. In physical systems self-similarity inevitably breaks down after some number of iterations (e.g. non-isothermal physics in molecular clouds at high densities), setting the dynamic range in which the model is applicable.}
\label{fig:fragm_model}
\end {center}
\end{figure}

Imagine an initial \myquote{cloud} of mass $M_0$ and size $R_0$ (e.g. for stars and cores this would be a GMC). This and all subsequently forming clouds are contracting and have a small, but finite chance $\epsilon$ of collapsing to infinite density and zero size (forming a star). Alternatively (with probability $1-\epsilon$) it fragments into a number of fragments ($1/\kappa$) with mass=$\kappa M_0$ after contracting by some factor $\lambda$ in size (see Fig. \ref{fig:fragm_model}). Afterwards the gas rearranges itself while conserving density. The process is repeated for each newly formed cloud fragment\footnote{Note that this process is highly hierarchical with multiple object forming out of a single cloud, making it different from the well-known single, spherical cloud evolution models \citep[e.g.][]{Larson_1969,Penston_1969}.}. To make our results normalizable we assume that there is a finite number of fragmentation events, in other words: the cascade is terminated. This is due to the breakdown of the scale-free assumption, in case of molecular clouds this is due to non-isothermal effects at high densities. Table \ref{tab:variables} shows the parameters and variables of the model\footnote{Note that the three parameters of the model all refer to mean quantities (e.g. $\kappa$ is the \emph{mean} relative mass of fragments). Our analysis aims to show that regardless of the underlying distributions, all self-similar fragmentation models produce statistically similar result.} . Note that to have an inertial range of significant size it must be true that $\epsilon\ll 1$.

\begin{table*}
	\centering
	\begin{tabular}{ | c | c |}
	\hline
  \textbf{Parameters} & {} \\
	\hline
   $\epsilon$ & Probability that a cloud does not fragment as it collapses. \\
	\hline
   $\lambda$ & Average contraction scale (of the parent) when fragmentation occurs.\\
	\hline
   $\kappa$ & Average mass of fragment relative to parent.\\
	\hline
	\hline
	\textbf{Variables} & {} \\
	\hline
	$M_n$ & Mass of $n^{\rm th}$ generation clouds. \\
	\hline
	$R_n$ & Size scale of $n^{\rm th}$ generation clouds. \\
	\hline
	$M_{*}(M_n)$ & Total mass of $n^{\rm th}$ generation stars. \\
	\hline
	$\rho_n$ & Initial density of $n^{\rm th}$ generation clouds. \\
	\hline
	$M_{\rm surv,n}$ & Mass of all surviving (non-collapsed) objects after $n$ fragmentation events. \\
	\hline
	\hline
	\textbf{Initial Conditions} & {} \\
	\hline
   $M_0$ & Mass of the initial cloud \\
	\hline
   $R_0$ & Size scale of the initial cloud.\\
	\hline
	\end{tabular}
	\caption{Parameters and variables in the toy fragmentation model we use to demonstrate the effects of scale-free behavior (see Fig. \ref{fig:fragm_model}).}
	\label{tab:variables}
\end{table*}

\subsection{Effects of Gravitational Collapse}\label{sec:Jeans}

Let us assume that fragmentation happens due to the Jeans instability (equivalently we could say all fragments have the same virial parameter). In an isothermal medium $M_{\rm Jeans}={\rm const.}\times \rho^{-1/2}$ so if we have a cloud that is marginally Jeans unstable ($M=M_{\rm Jeans}(1)$) then after it shrinks to $\lambda$ times its original size the new Jeans mass becomes
\be
M_{\rm Jeans}(\lambda)=M_{\rm Jeans}(1)\lambda^{3/2}.
\ee
This cloud then fragments into several pieces, each is roughly the Jeans mass, thus the number of fragments is
\be
N\equiv\kappa^{-1}=\frac{M}{M_{\rm Jeans}(\lambda)}=\lambda^{-3/2}.
\ee
In this case there is a clear connection between $\kappa$ and $\lambda$ such that
\be
\label{eq:jeans_kappa}
\ln \kappa=3/2 \ln \lambda.
\ee

Note that this simplistic analysis neglects other forms of cloud support (e.g. turbulence, rotation). Nevertheless, the simulations of \correlationpaper find that turbulence based fragmentation models yield results consistent with Eq. \ref{eq:jeans_kappa} (see Section \ref{sec:corr}). We will return to the importance of Eq. \ref{eq:jeans_kappa} below.

\section{Universal Scaling Laws}\label{sec:results}

For the case of scale-free fragmentation we can use our toy model to calculate the values of the variables from Table \ref{tab:variables} with relative ease. The results are shown in Table \ref{tab:results}.

\begin{table*}
	\centering
	\begin{tabular}{ | c | c | c | c | c | c|}
	\hline
  Generation ($n$) & $M_n/M_0$ & $R_n/R_0$ & $M_{*}(M_n)/M_0$ & $\rho_n/\rho_0$ & $M_{\rm surv,n}/M_0$\\
	\hline
   $0$ & $1$ & $1$ & $\epsilon$ & $1$ & 1\\
	\hline
   $1$ & $(1-\epsilon)\kappa$ & $\lambda [\kappa(1-\epsilon)]^{1/3}$ & $\epsilon (1-\epsilon)$ & $\lambda^{-3}$ & $(1-\epsilon)$\\
	\hline
   $2$ & $(1-\epsilon)^2\kappa^2$ & $\lambda^2 [\kappa(1-\epsilon)]^{2/3}$ & $\epsilon (1-\epsilon)^2$ & $\lambda^{-6}$ & $(1-\epsilon)^2$\\
	\hline
   $n$ & $(1-\epsilon)^n\kappa^n$ & $\lambda^n [\kappa(1-\epsilon)]^{n/3}$ & $\epsilon (1-\epsilon)^n$ & $\lambda^{-3 n}$ & $(1-\epsilon)^n$\\
	\hline
	\end{tabular}
	\caption{Values of different variables (see Table \ref{tab:variables} for definitions) for objects of different generations in our toy fragmentation model (see Fig. \ref{fig:fragm_model}). Having a large dynamic range implies $\epsilon\ll 1$ (otherwise all the mass would be at the largest scales) which simplifies most of these expressions. }
	\label{tab:results}
\end{table*}


\subsection{Mass Function}\label{sec:IMF}

First, if we look at the total mass of final objects (e.g. stars) in a given logarithmic mass bin ($M_{*}(M_n)$) in Table \ref{tab:results}, we find it to be proportional to $\epsilon (1-\epsilon)^n$. In realistic cases $\epsilon\ll 1$ (required to have a large dynamic range) so we get
\be
\frac{M_{*}(M_n)}{M_0}=\epsilon\left(\frac{M_n}{M_0}\right)^{\frac{\ln(1-\epsilon)}{\ln(1-\epsilon)+\ln\kappa}}\approx \epsilon\left(\frac{M_n}{M_0}\right)^{\frac{-\epsilon}{\ln\kappa}}\approx\epsilon=\mathrm{const,}
\ee
where we used that $n=\frac{\ln M_n/M_0}{\ln\left(1-\epsilon\right)+\ln\kappa}$ in the first equality which we can infer from Table \ref{tab:results}. The last approximation is only valid while $n\ll\epsilon^{-1}$, after that the expression becomes a very weak power-law (slope of $-\epsilon/\ln\kappa \approx 0$).

\begin{figure*}
\begin {center}
\includegraphics[width=0.33 \linewidth]{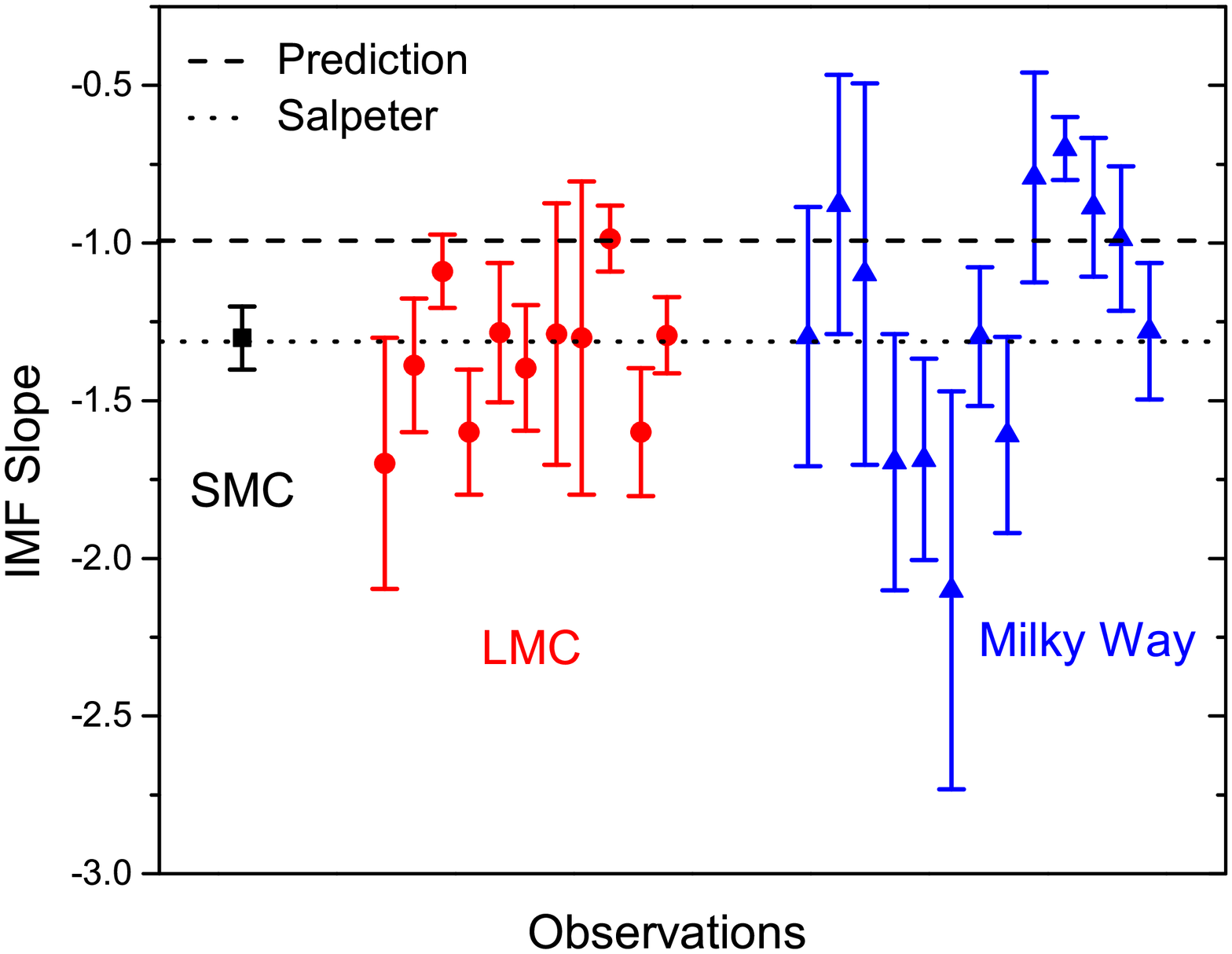}
\includegraphics[width=0.33 \linewidth]{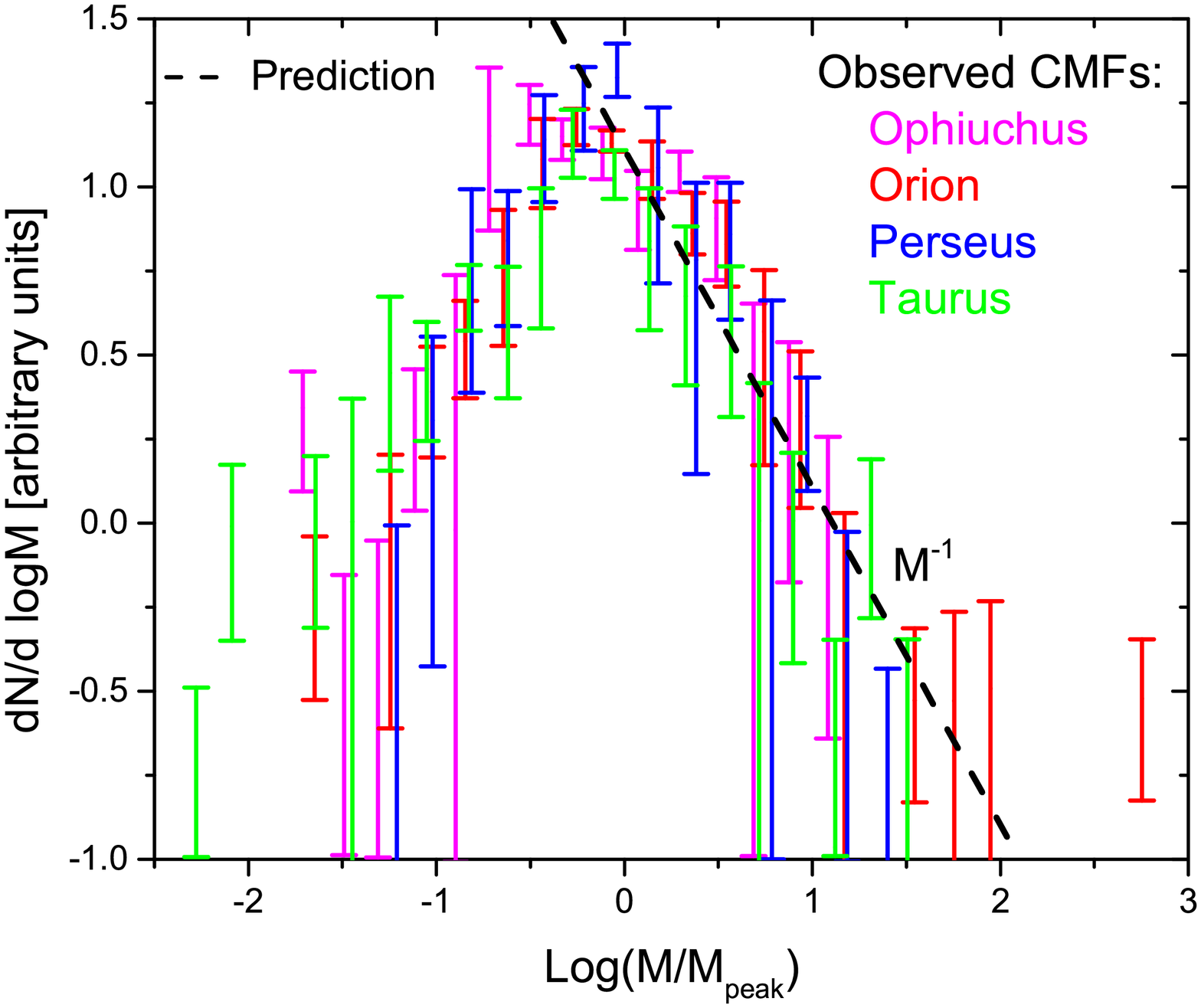}
\includegraphics[width=0.33 \linewidth]{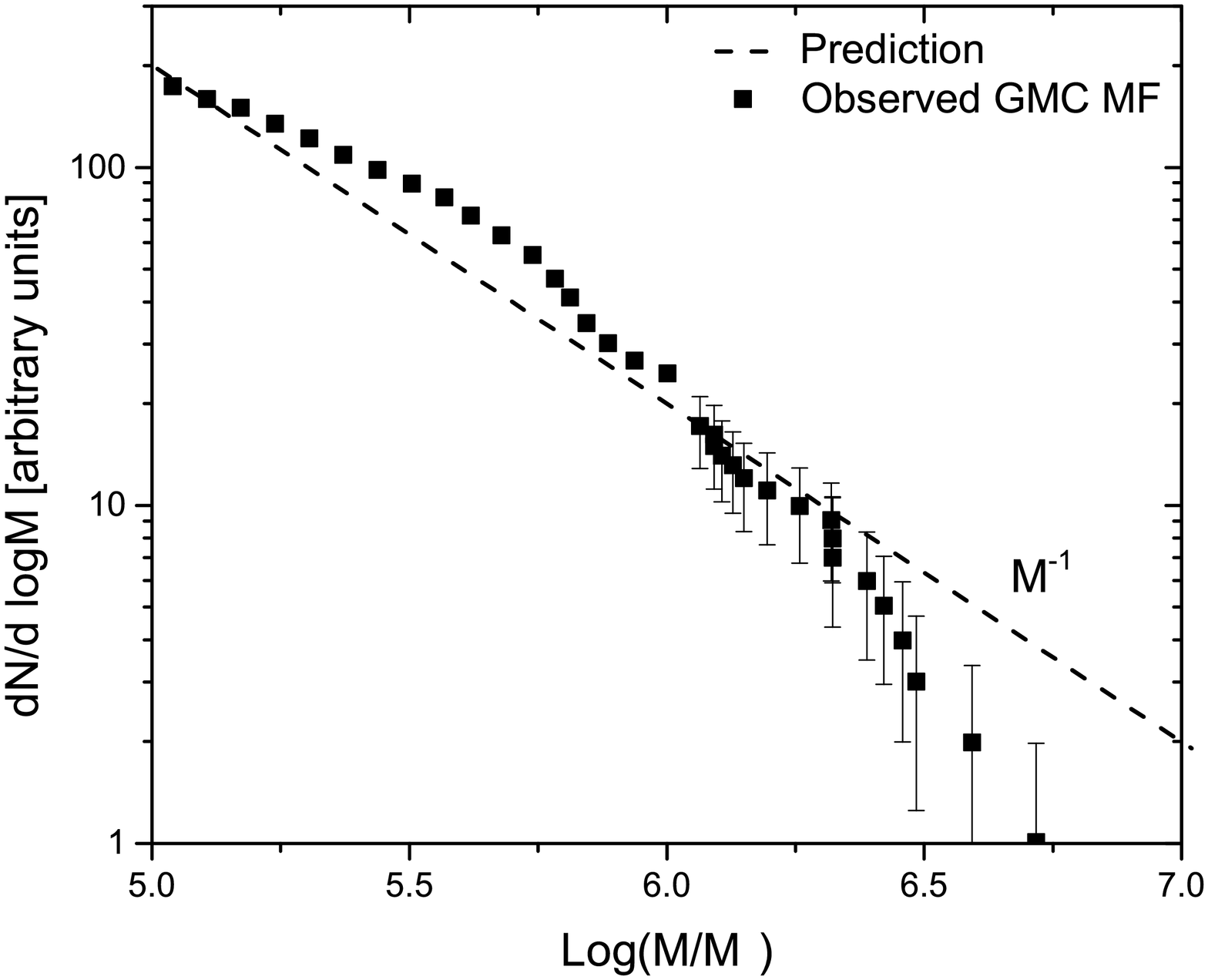}\\
\caption{
{\bf Left:} The observed slopes of the high mass end of the IMF \protect\citep{Massey_2003_IMF_slope} along with the canonical Salpeter slope (\protect\citealt{Salpeter_slope}, dotted line) and the prediction from our scale-free model (dashed line). Note that the error bars only account for fitting errors thus one should consider them lower estimates. The scale-free prediction of $M^{-1}$ is slightly shallower than the best fit slope of -1.35 \protect\citep{Salpeter_slope}. {\bf Middle:} The observed CMF in different regions \protect\citep{Sadavoy_observed_CMF} normalized in both axes. The observed high mass slope is roughly consistent with our prediction of $M^{-1}$ (dashed). {\bf Right:} The observed GMC mass function \protect\citep{Rosolowsky_2005_GMC} along with our prediction (dashed line). The observations are roughly in line with the scale-free predictions for scales below the high mass cutoff.}
\label{fig:mass_functions}
\end {center}
\end{figure*}

There is an equal amount of total mass per object mass in structures per logarithmic interval in mass of the final objects. Since the number of objects is (\textit{Mass per bin})/(\textit{Mass of an individual object}), this leads to a mass function of $\propto M^{-2}$. This is in rough agreement with the slopes of the IMF, the core, the GMC, the star cluster and the dark matter halo mass functions (\citealt{Alves_CMF_IMF_obs,IMF_universality,Rosolowsky_2005_GMC,Bik_2003_cluster_IMF,Warren_2006_DM_MF} respectively, see Fig. \ref{fig:mass_functions} for examples). Note that this conclusion is \emph{independent} of the model parameters $\kappa$, $\lambda$ and $\epsilon$ so long as there is a large dynamic range ($\epsilon\ll 1$) .

\subsection{Correlation Function}\label{sec:corr}

\begin{figure*}
\begin {center}
\includegraphics[width=0.45 \linewidth]{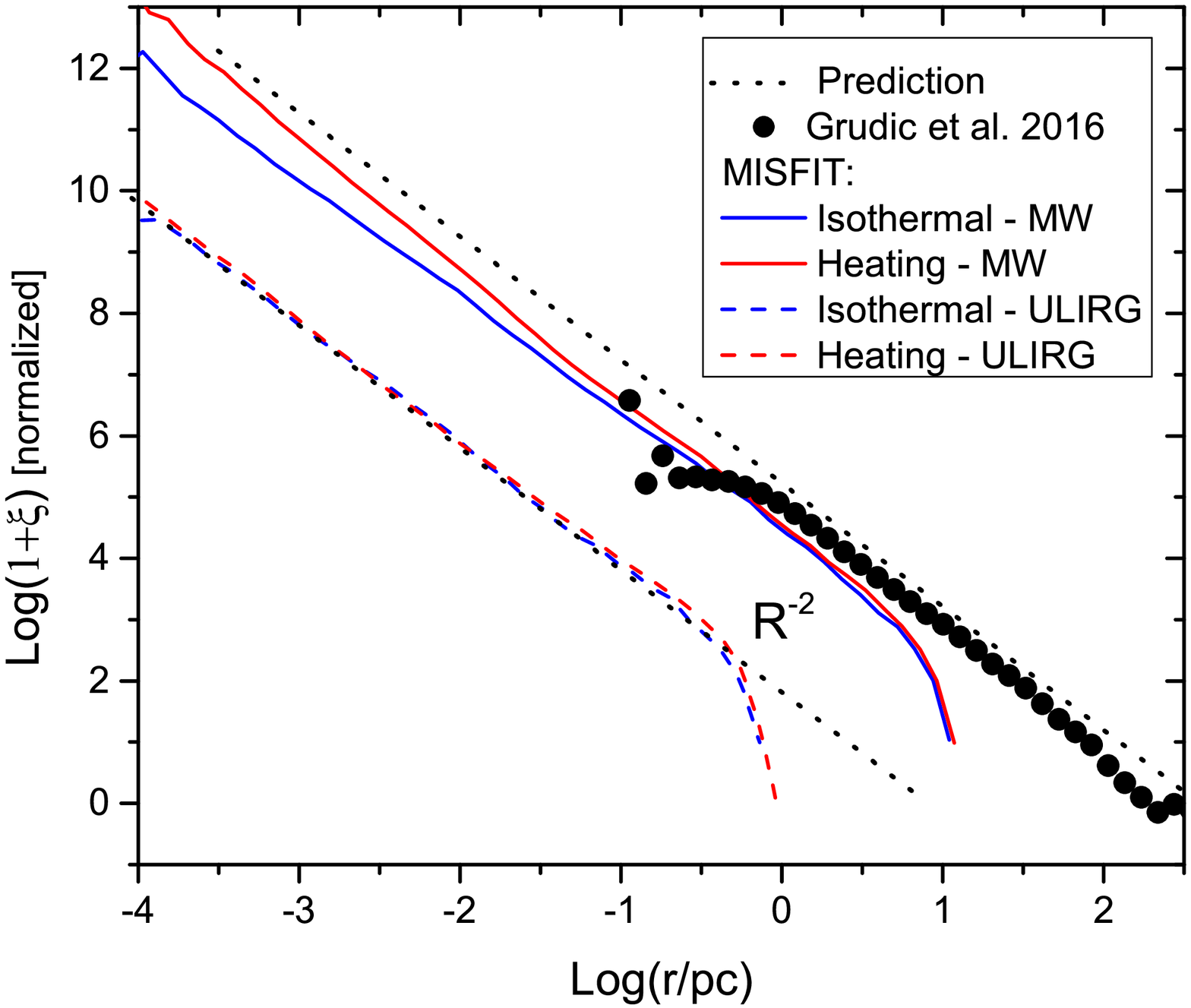}
\includegraphics[width=0.45 \linewidth]{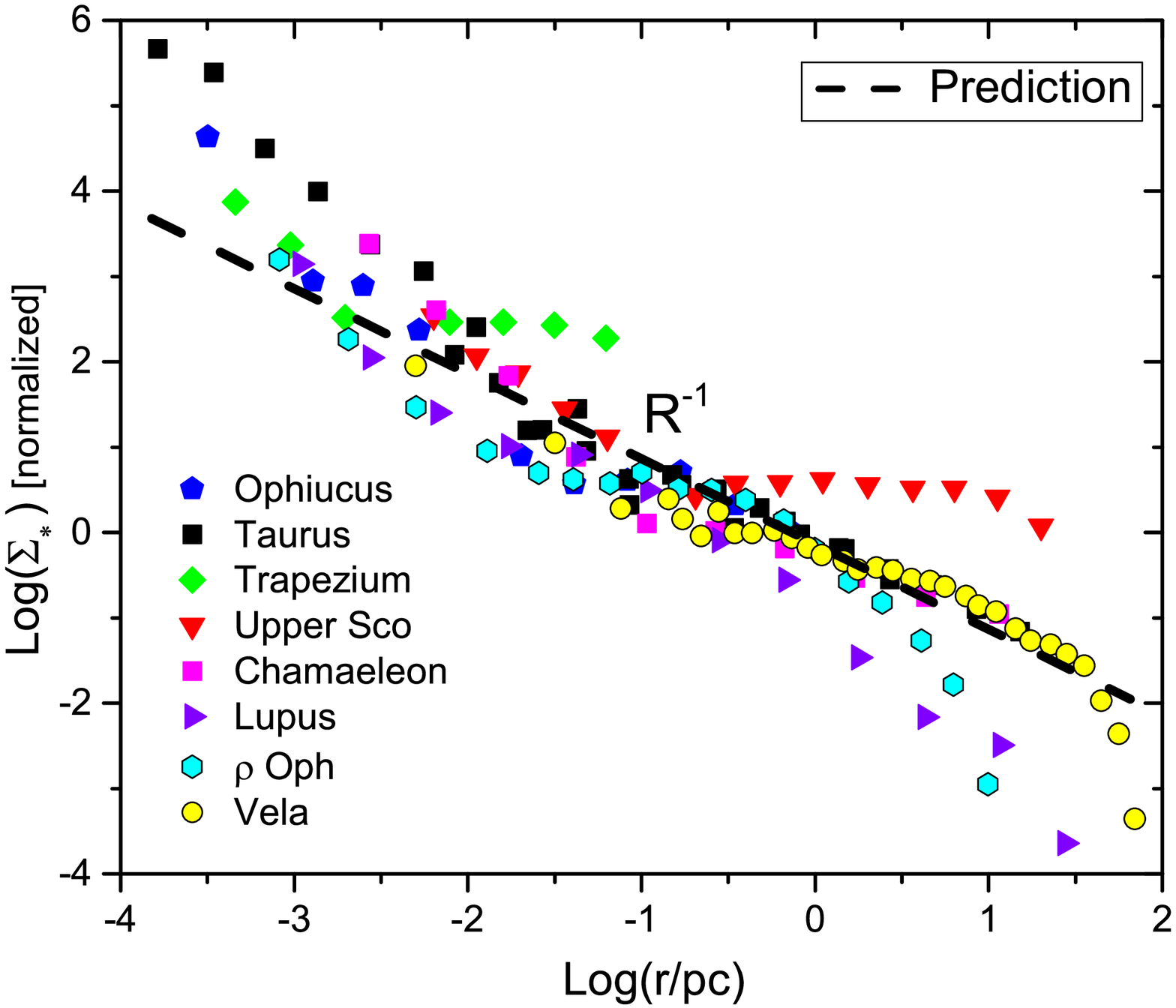}\\
\caption{
{\bf Left:} Stellar correlation function for different detailed semi-analytic star formation models in \correlationpaper , specifically an \myquote{isothermal} simulation and a model including \myquote{protostellar heating}, with MW-like and ULIRG-like initial conditions, along with the results of the full-numerical MHD simulations of \protect\cite{Grudic_2016}. The initial conditions and underlying physics have limited effect on the behavior, which is close to the predicted power law of -2. The different large scale cutoffs are introduced by the different initial cloud sizes, and the different normalization simply results from the different linewidth-size relation between the Milky Way and ULIRG cases. {\bf Right:} Observed surface density of neighboring stars ($\Sigma_*$, which is proportional to the projected correlation function $\xi_{\rm 2D}$) observed in different regions (labeled) Chamaeleon, Ophiucus, $\rho$ Oph, Taurus, Trapezium, Upper Sco, Lupus and Vela by \protect\citealt{Simon1997, Nakajima1998, Hartmann2002, Hennekemper2008, Kraus2008}. Our prediction of a power law with slope of -1 seems to match these on larger scales.}
\label{fig:allstar_corr_comp}
\end {center}
\end{figure*}

Let us now look at the correlation between objects of the same generation (mass). By only taking objects that formed after exactly $n$ fragmentation events we can calculate the fractal dimension of this ensemble. The number of such objects is $N_n=M_{*}(M_n)/M_n=\epsilon \kappa^{-n}$. If we focus on one of these objects and draw a sphere of radius $R_m$ around it we have $N_n(R_m)=\epsilon \kappa^{m-n}$ objects in it\footnote{This can be verified by considering that within $R_m$ radius of such an object are all other object that formed out of a single ancestor of $R_m$ size.}. Using $R_m=\lambda^m\left[\kappa(1-\epsilon)\right]^{m/3}$ from Table \ref{tab:results} we find the fractal dimension to be
\be
\label{eq:corr_slope}
D\sim \frac{\dderiv \ln{N_n(R_m)}}{\dderiv \ln{R_m}}=\frac{\ln \kappa}{\frac{1}{3}\ln \kappa+\frac{1}{3}\ln{(1-\epsilon)}+\ln \lambda}.
\ee 
Combined with Eq. \ref{eq:jeans_kappa} this yields $D=1$. Since our model is isotropic the fractal dimension is related to the the two-point correlation function. For the 3D and the (observable) 2D correlation functions this leads to $\xi_{\rm 3D}\propto r^{-2}$ and $\xi_{\rm 2D}\propto r^{-1}$ respectively (using Eq. \ref{eq:corr_d_general}, see Appendix \ref{sec:fractal} for details), which are in agreement with the simulation results from \correlationpaper and \cite{Grudic_2016}. These predictions also roughly agree with the observed stellar and DM halo correlation functions (see Fig. \ref{fig:allstar_corr_comp}) on intermediate scales\footnote{Note that due to the finite age of the Universe the spatial structure of DM on very large scales reflects the primordial density fluctuations and is not related to the subject of this paper.}. This is compared to simulations in Fig. \ref{fig:allstar_corr_comp}.

\subsection{Density PDF}\label{sec:density}

Using Table \ref{tab:results} we find the volume occupied by $n^{\rm th}$ generation objects at their formation ($V_n$) to be:
\be
\frac{V_n}{V_0}=\frac{M_{\rm surv,n}/\rho_n }{M_0/\rho_0}=\frac{M_{\rm surv,n}}{M_0}\left(\frac{\rho_n}{\rho_0}\right)^{-1}= (1-\epsilon)^n\left(\frac{\rho_n}{\rho_0}\right)^{-1}.
\ee
Using $n=-\frac{1}{3\ln\lambda}\ln(\rho_n/\rho_0)$ we can replace $n$, thus
\be
\frac{V_n}{V_0}=\exp{\left[-\ln(\rho_n/\rho_0)\ln(1-\epsilon)/(3 \ln \lambda)\right]}\left(\frac{\rho_n}{\rho_0}\right)^{-1}= \left(\frac{\rho_n}{\rho_0}\right)^{-1-\frac{\ln(1-\epsilon)}{3 \ln \lambda}}.
\ee
This fraction of the total volume once had objects of $rho_n$ density inside them (these either fragmented or collapsed), which means that these regions have an average density of $rho_n$. The binning by $n$ is a logarithmic so
\be 
V_n=\frac{\dderiv V}{\dderiv n}=\frac{\dderiv V}{\dderiv \ln \rho}\frac{\dderiv \ln \rho}{\dderiv n}\propto \frac{\dderiv V}{\dderiv \ln \rho}.
\ee
Thus the volume density PDF $\left(\frac{\dderiv V}{\dderiv \ln \rho}\right)$ should scale as
\be
\frac{\dderiv V}{\dderiv \ln \rho}\propto \rho^{-1-\frac{\ln(1-\epsilon)}{3\ln \lambda}}\approx \rho^{-1},
\ee
where we assumed $\epsilon\ll 1$ in the last step. Note that $\epsilon<1$ and $\lambda<1$ so the slope of the PDF is, in general, predicted to be somewhat steeper than -1. However, the approximate slope is, once again, independent of the model details. 

Note that this is a prediction for the density PDF of {\em all material which undergoes complete fragmentation} (e.g.\ goes on to form stars). It is not the same as the density PDF one would see at a given instant in time. To calculate the latter (the observable PDF), we need to convolve the PDF of clouds as they collapse with some observable ``lifetime'' (our model, thus far, makes no assumptions about the amount of {\em time} each step in the process actually takes). Since this requires some outside assumptions (which are unconstrained by the fundamental nature of fragmentation, and could be related to e.g.\ cooling, or dynamical, or turbulent processes), we do not wish to argue for any particular model for the lifetime in this paper, and so should take the comparison with observations with some caution. A reasonable possibility, however, would be to take the lifetime to be proportional to the freefall time $t_{\rm freefall}\sim \rho^{-1/2}$; this would steepen the proposed slope by $1/2$ (giving an observable slope of $-3/2$). Other assumptions involving ``slower'' collapse (longer lifetimes) will generally produce slopes between $-1$ and $-3/2$. 

Another important effect comes from the the density profile of the individual clouds. In our model we assumed all clouds to be homogeneous while in reality they develop significant density gradients. The overall density PDF is a convolution of this density profile and the PDF we predicted for homogeneous clouds.

\subsubsection{Previous Results in the Literature}

There have been previous significant theoretical efforts to model the slope of the density PDF. Many of these were based on numerical simulations \citep[e.g.][]{scalo:1998.turb.density.pdf,ostriker:1999.density.pdf,klessen:2001.sf.cloud.pdf,vazquez-semadeni:2001.nh.pdf.gmc,audit:2010.gmc.massfunctions,federrath:2010.obs.vs.sim.turb.compare,ballesteros-paredes:2011.dens.pdf.vs.selfgrav,federrath.2015:density.pdf.and.sfr.in.polytropic.turbulence,squire.hopkins:turb.density.pdf}, which we discuss below. There have also been several analytic models proposed, many of which are similar in spirit to \citet{Girichidis_density_PDF_2014} who assumed self-similar collapse, with individual clouds observable (or ``surviving'') at a given density for a time proportional to their free-fall time, and predict a slope of $-1.54$ (while our model predicts $-1.5$, for the same observable-time assumption).\footnote{Note that there is a small error in Eq.~12 in \citet{Girichidis_density_PDF_2014} where the authors inadvertently assumed that volume is conserved in cloud evolution, despite modeling shrinking clouds. This can be easily corrected by replacing their Eq.~12 with the mass-conserving version of the equation (which they present earlier); after accounting for this correction (which amounts to one power of $\rho$) the result is that their $-1.54$ result is directly comparable to our $-1.5$.}

\subsubsection{Column Density PDF}\label{sec:column_density}

Because the volume density PDF itself is not directly observable, let us calculate the PDF for the line integrated (surface) density $\Sigma$. To do that we choose a random line of sight to integrate along that goes through the cloud we are interested in. Let us denote the chance that such a random line goes through one of the dense substructures of the cloud with $p$ ($\epsilon\ll 1$ so we neglect the case when cloud does not have substructure). If the line avoids the substructures the line integrated density is $\Sigma_0\sim\rho_0 R_0$ whereas if it hits the dense region we get $\Sigma_1\sim(R_0-R_1)\rho_0+R_1 \rho_1$. From Table \ref{tab:results} it is easy to see that
\be
\frac{\Sigma_1}{\Sigma_0}\sim1+\frac{R_1}{R_0}\left(\frac{\rho_1}{\rho_0}-1\right)=1+\lambda [\kappa (1-\epsilon)]^{1/3}\left(\lambda^{-3}-1\right)\approx 1+\lambda^{-2}\kappa^{1/3}.
\ee
Using the Jeans collapse condition (Eq. \ref{eq:jeans_kappa}) we find that $\lambda^{-2}\kappa^{1/3}=\lambda^{-3/2}$ which is much greater than 1 for realistic cases. This means that the line integrated density is dominated by the densest substructure along the line of sight. In general we get:
\be 
\frac{\Sigma_n}{\Sigma_0}=\left(\lambda^{-2}\kappa^{1/3}\right)^n.
\label{eq:sigma_n}
\ee
Since $p$ is the probability of hitting the dense substructure of a cloud, the probability of the densest region along our line of sight to be from generation $n$ is $P_n=(1-p)p^n$ as it needs to penetrate exactly $n$ levels of substructure. We can directly calculate $p$ because it is the cross section of the dense subregions relative to their parent (while taking into account that there are $\kappa^{-1}$ of them), so
\be
p=\frac{\kappa^{-1}\pi R_{n+1}^2}{\pi R_n^2}=\kappa^{-1}[\kappa(1-\epsilon)]^{2/3}\lambda^2\approx \kappa^{-1/3}\lambda^2.
\label{eq:prob_value}
\ee
Now, using Eq. \ref{eq:prob_value} and Eq. \ref{eq:sigma_n} we find the total area with $\Sigma_n$ surface density is
\be 
\frac{A_n}{A_0}=(1-p)p^n\propto p^n=\left(\frac{\Sigma_n}{\Sigma_0}\right)^{-\frac{1}{n}n}\propto\Sigma^{-1}.
\ee
Similar to the volume density case the logarithmic binning in $n$ leads to
\be 
A_n=\frac{\dderiv A}{\dderiv n}=\frac{\dderiv A}{\dderiv \ln \Sigma}\frac{\dderiv \ln \Sigma}{\dderiv n}\propto \frac{\dderiv A}{\dderiv \ln \Sigma} \rightarrow\frac{\dderiv A}{\dderiv \ln \Sigma}\propto \Sigma^{-1},
\label{eq:column_density_distrib}
\ee
where we have used Eq.\ref{eq:sigma_n}.

Just like the volume density PDF, the surface density PDF is affected by the finite observable lifetimes of clouds; as noted before this will steepen the slope, most likely producing final slopes roughly between $-1$ and $-2$. So any comparison with observations must be considered with caution here.

\begin{figure}
\begin {center}
\includegraphics[width=\linewidth]{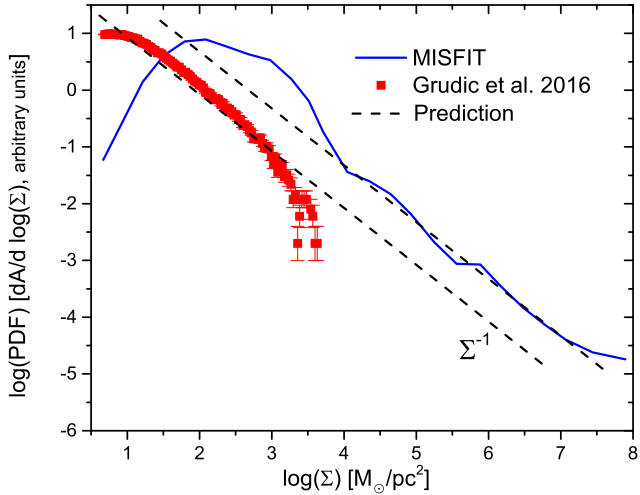}
\caption{Surface density ($\Sigma$) PDF of star-forming gas in the detailed MHD simulation of \protect\cite{Grudic_2016} and in one of the GMC collapse simulations using the MISFIT semi-analytical framework (see \protect\citealt{guszejnov_GMC_IMF} for details). After the fragmentation cascade begins the system develops a $\Sigma^{-1}$ power law tail in line with the predictions of the scale-free model. Note that both of PDFs take only the star forming gas into account.}
\label{fig:sim}
\end {center}
\end{figure}

\begin{figure}
\begin {center}
\includegraphics[width=\linewidth]{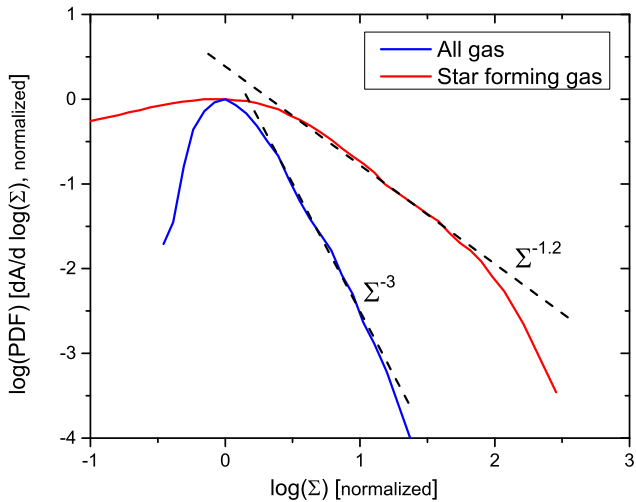}
\caption{Surface density ($\Sigma$) PDF of star-forming gas vs all gas associated with the molecular cloud in the detailed MHD simulation of \protect\cite{Grudic_2016}. We define star forming gas as fluid elements in a converging flow within a self-gravitating structure of molecular gas. Both axes are normalized so that the peak of both PDFs are unity at $\Sigma=1$. Similar to \protect\cite{Lin_cloud_structure_2017} we find that star forming gas produces a surface density PDF with a power-law slope of -1.2, close to our prediction of -1, while the distribution for the total gas has a much steeper power-law tail. This is because we integrate along the entire line of sight, thus our measurement is sensitive to the background density profile of the non-star-forming gas.}
\label{fig:sf_vs_all}
\end {center}
\end{figure}

Fig. \ref{fig:sim} shows surface density PDFs in two simulations: the MISFIT semi-analytic framework \citep[see][]{guszejnov_GMC_IMF} and the detailed MHD simulations of \cite{Grudic_2016}. In both simulations star forming regions develop a similar power-law tail once the fragmentation cascade begins, a phenomenon that has been observed in other simulations (e.g. \citealt{Kritsuk_dens_lognormal}) as well. Here (in the simulations) we have the advantage that we can {\em specifically} isolate gas which is un-ambiguously known to be star-forming, which also means it is self-gravitating and undergoing fragmentation (thus, is in the regime where our model should apply). In Fig. \ref{fig:sf_vs_all} we show that this can have drastic effects. The star forming regions we are observing are embedded in much larger reservoirs of gas which is not undergoing a fragmentation cascade, so our model is not applicable there\footnote{This does not mean that this larger reservoir is not evolving, it roughly follows the isothermal collapse models formulated for spherically symmetric, non-fragmenting clouds, which leads to the development of its own density profile and PDF. The key difference is that in this regime pressure effects are not negligible. For a discussion of the resulting density PDF see \cite{Kritsuk_dens_lognormal}.}. Meanwhile the line of sight for our observation integrates the density in these regions too. The net result is that instead of the PDF of the star forming region we see a convolution of that and the background density profile, which leads to a much steeper density PDF than predicted by our model.

Comparing to observations is more difficult because of both the lifetime effects and the difficulty of separating ``non-fragmenting'' (even if dense) cloud material from that which is truly experiencing runaway fragmentation. 
But there have been a number of studies of the density structure of molecular clouds
(e.g. \citet{Kainulainen_power_law_tail,Schneider_2013_orionB_density,Lombardi_2014_column_density}) which found that the column density PDF in molecular clouds is best described by a lognormal peak at low and a power law tail at high densities. On average these studies have found an average slope of $\sim\,-2.3$ (significantly steeper than our analytic prediction). But it has been shown \citep[see][]{Schneider_2015} that these measurements are actually dominated by the mass profile of the parent clouds\footnote{For a spherically-symmetric cloud with a radial mass profile of $\rho \propto r^{-\beta}$, one obtains a volume-density PDF of ${\dderiv V}/{\dderiv \ln \rho} \propto \rho^{-3/\beta}$ and a surface-density PDF of ${\dderiv A}/{\dderiv \ln \Sigma} \propto \Sigma^{-2/(\beta-1)}$, so a slope of $\sim -2$ in ${\dderiv A}/{\dderiv \ln \Sigma}$ corresponds to an isothermal-sphere density profile $\rho\propto r^{-2}$.}. In other words, just like in the simulation, the measured surface density PDF is a convolution of the global mass profile, and the PDF driven by fragmentation and turbulence within an annulus of constant density. Our result in Eq. \ref{eq:column_density_distrib} is the surface density PDF that would be measured in such a fragmenting, constant background density annulus of a cloud -- we intentionally have {\em not} made any assumption about embedding our (local) fragmentation model inside some (global) mass profile of a parent cloud (if we did, we could easily fit these observations, but it would only reflect the assumed mass profile). 

Allowing for a background density profile with locally collapsing regions, one generically expects the following: at the highest densities, the density PDF should be dominated by collapsing, star-forming regions, which should (if our model is correct) follow our prediction with an approximately $-1$ slope. At lower (intermediate) densities, where not all the material is locally self-gravitating and fragmenting, one becomes dominated by the combination of turbulent density fluctuations and the background density profile of the cloud, and the PDF will have a steeper slope that matches the cloud profile. Interestingly, \citet{Schneider_two_power_law_2015} claim to see almost exactly such a transition, with steeper slopes $\sim -2.3$ at intermediate densities (matching their fits to the global mass profile) and a shallower slope (or ``excess'' in their terms) appearing at approximately $\Sigma \gtrsim 100\,\langle \Sigma \rangle$ with a slope $\approx -1 \pm 0.2$. Similarly, \citet{Lin_cloud_structure_2017} see in the survey of clouds that the PDF becomes systematically shallower, approaching $-1$, as clouds (or cloud regions) become more actively star-forming. We should also note that similar results have been seen in other numerical simulations \citep[e.g.][]{Burkhart_cloud_sim_2015}.

\subsection{Cluster Mass Profile}\label{sec:cluster}

Let us assume that some fraction of objects formed remain gravitationally bound to each other. We expect that the clustered substructures that formed from fragmentation will eventually merge together into a cluster with a density profile that decreases monotonically. Let us derive the power law index of this profile.

Using our model we can calculate the relation between the densities and the survivor masses before the objects rearrange themselves into clusters. For density, take their at-formation value ($\rho_n$). We can express the index $n$ as:
\begin{eqnarray}
n=\frac{\ln(M_{\rm surv,n}/M_0)}{\ln(1-\epsilon)},\\
n=-\frac{1}{3\ln\lambda}\ln(\rho_n/\rho_0),
\end{eqnarray}
which leads to
\be
\frac{\dderiv \ln(M_{\rm surv,n}/M_0)}{\dderiv \ln(\rho_n/\rho_0)}=-\frac{\ln(1-\epsilon)}{3\ln\lambda}={\rm const.}\approx 0
\label{eq:cluster_1}
\ee
Let us assume (for now) that after formation the objects rearrange themselves to form clusters, while preserving the local volume density - i.e. the local density (of stars) around a star does not change dramatically before/after the re-arrangement. This is motivated by the fact that during mergers, tidal shredding of an object with density $\rho$ occurs at an orbital radius $R$, where the mean density enclosed within $R$ is approximately $\rho$. This means that the amount of mass at different density levels must be the same as before the rearrangement, thus
\be
\frac{\dderiv \ln(M_{\rm cl}/M_{\rm cl, 0})}{\dderiv \ln(\rho_{\rm cl}/\rho_{\rm cl,0})}=\frac{\dderiv \ln(M_{\rm surv,n}/M_0)}{\dderiv \ln(\rho_n/\rho_0)}.
\label{eq:phase_dens_conserv}
\ee
Let us assume the relaxed cluster has a power-law density profile: $\rho_{\rm cl}\propto R^{\beta-3}$. Also, $\epsilon\ll 1$ thus the right hand side of Eq. \ref{eq:cluster_1} is zero. This leads to
\be
\frac{\beta}{\beta-3}=0,
\ee
thus $\beta=0$, so the mass profile of a bound cluster that results from the assembly of substructures formed in a scale-free fragmentation cascade is $\rho_{\rm cl}\propto R^{-3}$.

We can repeat the same exercise while assuming that \emph{phase space density} ($\rho_p$) is conserved instead of real space density, as per Liouville's theorem. This is only true however if $\rho_p$ is resolved on infinitely fine scales, as elements of higher phase space density effectively get stretched out and diluted in phase space so that the final observed coarse-grained phase space density is generally lesser  than the initial \citep{Lynden_Bell_1967}. However, we may still suppose that our self-similarity condition means that the evolution operator on the coarse-grained phase space density can only map an initially flat ($\dderiv M/\dderiv \ln\rho_p \sim 0$) phase space distribution into another flat distribution.

We can approximate the phase space density as $\rho_p\approx \frac{\rho}{\sigma^3}$ where $\sigma$ is the velocity dispersion. Assuming that the collapsing clouds are virialized we can write
\be
\sigma_n^2=\frac{G M_n}{R_n}=\frac{G M_0}{R_0} (1-\epsilon)^n\kappa^n \lambda^{-n}[\kappa(1-\epsilon)]^{-n/3},
\ee
which leads to
\be
\rho_{p,n}=\rho_{p,0}\left(\lambda^{3/2}\kappa (1-\epsilon)\right)^{-n}.
\ee
From here we can formulate the surviving mass per phase density (similar to Eq. \ref{eq:cluster_1}):
\be
\frac{\dderiv \ln(M_{\rm surv,n}/M_0)}{\dderiv \ln(\rho_{p,n}/\rho_{p,0})}=-\frac{\ln(1-\epsilon)}{-\frac{3}{2}\ln \lambda-\ln\kappa-\ln(1-\epsilon)}\approx 0,
\label{eq:cluster_phase}
\ee
where, in the last step, we used the assumption that $\epsilon\ll 1$. Since we are interested in the asymptotic case at large radii. the mass enclosed is approximately converged ($M\approx{\rm const.}$), thus
\be
\sigma_{\rm cl}^2(R)=\frac{G M}{R}\propto R^{-1},
\ee
so we get
\be
\rho_{p,\rm cl}(R)\propto \frac{\frac{M_{\rm cl}(R)}{R^{3}}}{R^{-3/2}}\propto R^{\beta-3/2}.
\ee
After plugging into Eq. \ref{eq:phase_dens_conserv} and using Eq. \ref{eq:cluster_phase} this yields
\be
\frac{\beta}{\beta-3/2}=0,
\ee
which, once again, means that $\beta=0$ leading to $\rho_{\rm cl}\propto R^{-3}$.

\begin{figure}
\begin {center}
\includegraphics[width=\linewidth]{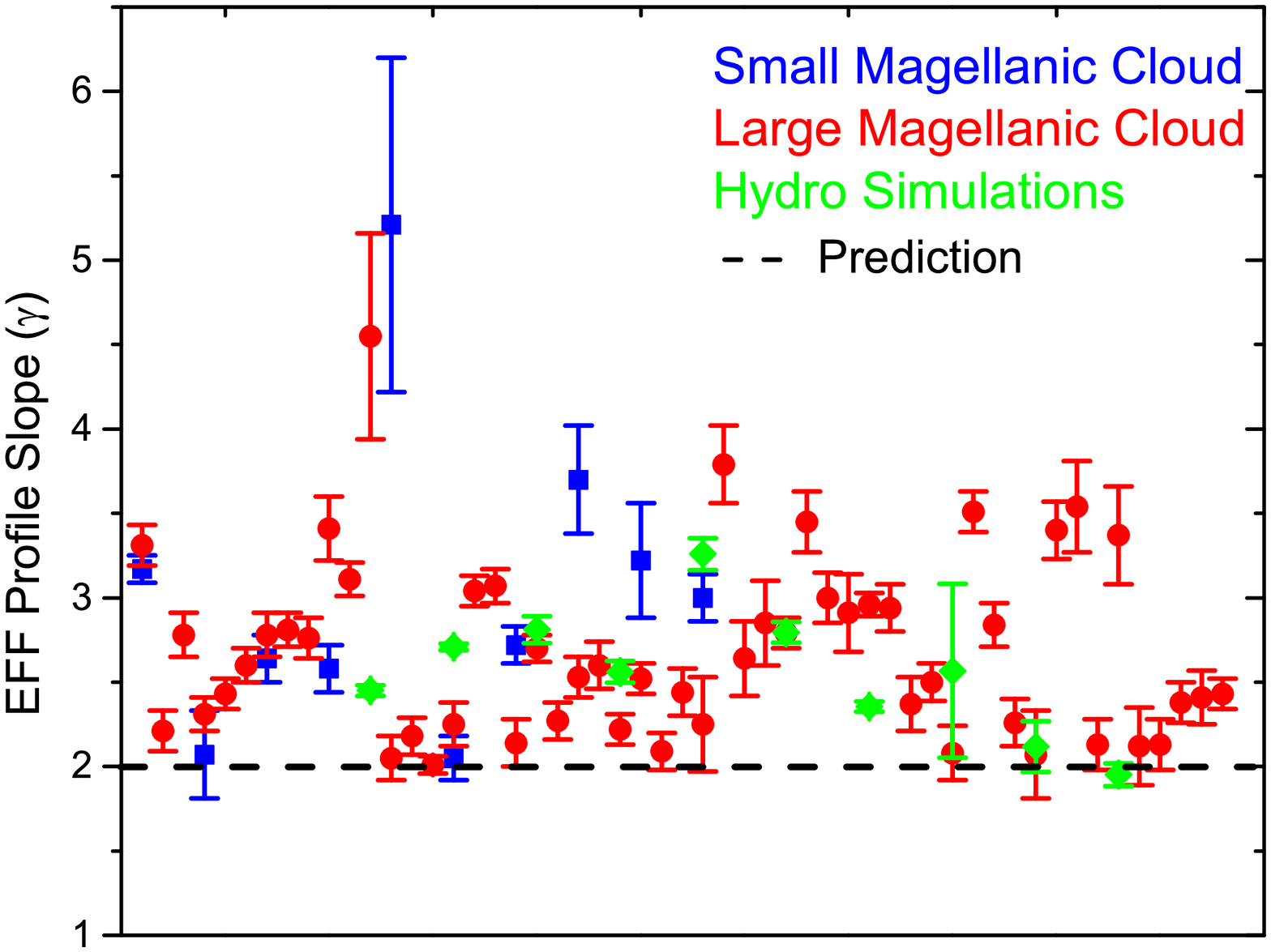}
\caption{The observed surface brightness profile slopes (using the profile of \citealt{Elson1987}) of several stellar clusters from the LMC \protect\citep{Mackey_Gilmore_2003} and the SMC \protect\citep{Mackey_Gilmore_2003_SMC} along with the scale-free prediction (dashed line). For reference we included the cluster population produced in one of the detailed MHD simulations of \protect\cite{Grudic_2016}. Both observed and simulated clusters seem to have steeper profiles than predicted by the scale-free model. This is due to the fact that $\gamma \sim 2$ is only the limiting case for an infinitely-extended hierarchical merger history (see \protect\citealt{grudic_2017} for more details).}
\label{fig:observed_cluster_gamma}
\end {center}
\end{figure}

Now, let us compare our prediction with observations. The observed brightness profile of young star clusters is often parametrized using the EFF profile (\citealt{Elson1987}):
\be
\mu\left(r\right) = \mu_{0} \left( 1 + \frac{r^2}{a^2}\right)^{-\gamma/2},
\label{eq:EFF_profile}
\ee
where $\mu_0$ is a constant, $a$ is the cluster scale radius and $\gamma$ is the power law index of the outer profile. Because $\mu\propto \Sigma\sim \rho R$ this represents an outer column density profile with $-(\gamma+1)$ slope, so in this parametrization our prediction is $\gamma=2$. Observations young massive clusters, both within the Local Group \citep{Elson1987,Mackey_Gilmore_2003,Mackey_Gilmore_2003_SMC,Portegies_young_clusters} and in extragalactic environments \citep{larsen_2004,ryon_2015_M83_clusters} have found that typically $\gamma\in[2,4]$, with a median around 2.5 (see Fig. \ref{fig:observed_cluster_gamma}). Meanwhile the density profile of dark matter halos is well fit by the NFW profile \cite{NFW_1996} that simplifies to a $\rho\propto r^{-3}$ on large scales, corresponding to $\gamma=2$. In \citet{grudic_2017} we explore the physics of hierarchical cluster assembly, and its imprint upon the density profiles of objects thus formed, in greater detail.

\subsection{Comparison with multi-physics simulations}

In Figures \ref{fig:allstar_corr_comp}, \ref{fig:sim} and \ref{fig:observed_cluster_gamma} we show that the predictions of our simple scale-free model are in good agreement with the simulations of \protect\cite{Grudic_2016} that follow the process of star formation in a molecular cloud. Our toy model of the fragmentation cascade and these simulations represent two extremes of approach to the problem of star formation: one seeks to minimize complexity, whereas the other seeks to maximize realism by including a variety of pieces of physics that could potentially be relevant. Under the forces of MHD and gravity the star-forming clouds collapse into a complicated structure of dense filaments \citep[e.g.][]{Collins_2012} that is difficult to reconcile with an idealized picture of fragmenting gas balls (e.g. Figure \ref{fig:fragm_model}). The realistic ISM equation of state with radiative cooling and stellar feedback (see \citealt{Hopkins_fire2} for details) leads to the formation of a multi-phase ISM, so the isothermality we have assumed does not strictly apply. 

Despite all of these complications the simulations and the toy model ultimately arrive at the same scaling relations. The simulation follows hierarchical fragmentation over a significant dynamic range in which the process, despite all the above complications, is roughly scale-free. Although real star-forming systems are more complicated than any simulation, this apparent robustness to such complications suggests that the observed scalings could be (to first order) explained as fundamental consequences of scale-free structure formation.

\section{Conclusions}\label{sec:consclusions}

In this paper we showed that there are universal scaling relations that generally arise in scale-free models of structure formation with a large but finite dynamic range and no correlation between scales. These relations are shared between very different phenomena, including the formation of stars, protostellar cores, clumps, giant molecular clouds, star clusters and even dark matter halos. Despite their differences all these processes can be approximately described  by the dimensionless version of the pressure-free Euler equation with self-gravity. The only parameter of this equation is the virial parameter, which is of course how we define gravitationally bound structures. Thus a hierarchical structure building process would follow the same equation for all these systems on a wide range of scales. This means that (to first order) the formation of these (very different) gravitationally bound structures produces the same scaling relations for a wide range of physical quantities.

We demonstrate these universal scaling relations for a subclass of processes that can be described as a fragmentation cascade (e.g. formation of molecular clouds). We present a minimalist model of self-similar fragmentation with which we show that any scale-free model with a large dynamic range where the scales are uncorrelated (the fluctuations on different scales \myquote{don't know about each other}) is able to reproduce the following scaling relations:
\begin{itemize}
\item \textit{Mass functions}: $\frac{\dderiv N}{\dderiv \ln M}\propto M^{-1}$. In the scale-free regime we expect objects of all sizes to follow this scaling relation. For the stellar initial mass function (\emph{IMF}, see \citealt{IMF_review,IMF_universality}), the protostellar core mass function (\emph{CMF}, see \citealt{Sadavoy_observed_CMF}), the \emph{molecular clump mass function} \citep[see][]{Johnstone_Bally_2006_clump_MF}, the giant molecular cloud mass function (\emph{GMC mass function}, see \citealt{Rosolowsky_2005_GMC}), the \emph{star cluster mass function} \citep[see][]{Bik_2003_cluster_IMF} and the dark matter halo mass function \citep{PressSchechter} this regime is observed over a wide dynamic range above/below some minimum/maximum scale where our assumptions are violated. This relation means that there is a comparable amount of mass in objects at all mass scales.
\item \textit{Density and column density PDF}: $\left(\frac{\dderiv V}{\dderiv \ln \rho}\right)_{\rm observ}\propto \rho^{-1}$ and $\frac{\dderiv A}{\dderiv \ln \Sigma}\propto \Sigma^{-1}$. The observed \emph{column density PDFs of star forming molecular clouds} exhibit a power-law tail consistent with this prediction \citep[see][]{Schneider_two_power_law_2015}, which is built up by the hierarchical fragmentation of clouds. This is a scale-free process as long as we are in the isothermal phase of the ISM. The scaling can be understood as having an equal amount of mass at all density scales. Note that our model does not account for observational biases (e.g. cloud lifetimes) which can lead to systematic differences with observations.
\item \textit{Correlation functions}: $\xi_{\rm 3D}\propto r^{-2}$ and $\xi_{\rm 2D}\propto r^{-1}$. Note that unlike the predictions above, this depends on the form on the fragmentation criterion so it is not completely generic. But if we assume that the criterion is set by gravity (as it is in all the above cases), this provides a good fit to the observed behavior of the \emph{stellar correlation function} at larger scales \citep[e.g.][]{Hennekemper2008,Kraus2008}. Similarly the observed \emph{protostellar core correlation function} is roughly consistent with this power-law behavior \citep[see][]{Stanke2006}. For the \emph{correlation function of star clusters}, this prediction agrees well with observations on scales smaller than the galactic scale height, above which the problem changes dimensionality \citep[see][]{grasha_2017_cluster_correlation}. For the case of DM halos this prediction is consistent with the measured correlation on intermediate scales \citep{Baugh_1996_DM_correlation,Soltan_2015_galaxy_correlation} as the finite age of the Universe causes the very large scale structures to reflect the primordial initial conditions. A possible interpretation of this slope is that it implies a fractal dimension of unity which further implies filamentary structures. Note that this is not generally true, for example our fragmentation model of Sec. \ref{sec_method} or even the Cantor-dust produces similar correlation functions, while none of them exhibit filamentary behavior.
\item \textit{Cluster mass profile}: $\rho_{\rm cl}\propto R^{-3}$ (or $\gamma=2$ using the EFF fitting function from \citealt{Elson1987}), if we further assume that the process responsible for arranging objects into clusters is also scale-free. This is a somewhat shallower slope than what is observed for star clusters \citep[see][]{Mackey_Gilmore_2003,Mackey_Gilmore_2003_SMC}, because $\gamma \sim 2$ is only the limiting case for an infinitely extended hierarchical merger history (see \citealt{Grudic_2016} for more details). This scaling can be understood as the cluster having equal mass at each distance scale from its centre, as a result of free mixing.
\end{itemize}
We wish to emphasize that these scaling relation are not unique to the fragmentation cascade paradigm. Any scale-free structure building process that satisfies the requirements of Sec. \ref{sec:universal_behaviour} would recover them. The reader should also note that these universal scaling relations can not explain all observed scaling laws. For example, the linewidth-size and mass-size relations in molecular gas ($\sigma^2\propto R$ and $M\propto R^2$ respectively, see \citealt{Larson_law,Scoville_1987_mass_linewidth_size,Bolatto_2008} for details) require additional physics. One possible candidate is supersonic turbulence, which naturally reproduces the linewidth-size relation \citep{Kritsuk_larson_supersonic_origin} due to its power-spectrum  \citep{Murray_supersonic_Burgers_analytic}. If one further assumes that the clouds are virialized, it naturally follows that $GM/R\sim \sigma^2$ so $M\propto R^2$.

It is easy to see that the arguments in this paper are invariant under time reversal transformation (in other words they don't rely on an \myquote{arrow of time}). For the toy model presented in Sec. \ref{sec_method}, this means that small objects merge to form ever bigger ones. This means that the predicted scalings should be present not only in hierarchical fragmentation but in the time-reversed process of hierarchical merging as well. This is the growth mechanism of galaxies and dark matter halos \citep{Lacey_Cole_1993_DM_merger,Kauffmann_1993_DM_galaxy_meger}.

Our simple arguments provide a generic, natural reason why so many different models, with seemingly very different physics, have been able to reproduce some or all of these scaling relations: the relations do \emph{not} depend on the detailed physics, so long as it is scale-free. It follows that these scaling relations can not be used to observationally differentiate theories (i.e. almost any theory can reproduce the IMF slope) because to first order they all give the same answers. One should instead test models against unrelated scaling laws (e.g. linewidth-size relation, see above) or investigate the physical scale where the model predicts the scale-free assumption to break down (e.g. IMF turnover) and compare them to observations in that regime.

\acknowledgments
We would like to thank Michael S. Fall and Christopher F. McKee for their invaluable comments.\\
Support for PFH, DG and MG was provided by an Alfred P. Sloan Research Fellowship, NASA ATP Grant NNX14AH35G, and NSF Collaborative Research Grant \#1411920 and CAREER grant \#1455342. Numerical calculations were run on the Caltech compute cluster ``Zwicky'' (NSF MRI award \#PHY-0960291) and allocation TG-AST130039 granted by the Extreme Science and Engineering Discovery Environment (XSEDE) supported by the NSF.

\bibliographystyle{mnras}
\bibliography{bibliography}

\begin{thebibliography}{}
\makeatletter
\relax
\def\mn@urlcharsother{\let\do\@makeother \do\$\do\&\do\#\do\^\do\_\do\%\do\~}
\def\mn@doi{\begingroup\mn@urlcharsother \@ifnextchar [ {\mn@doi@}
  {\mn@doi@[]}}
\def\mn@doi@[#1]#2{\def\@tempa{#1}\ifx\@tempa\@empty \href
  {http://dx.doi.org/#2} {doi:#2}\else \href {http://dx.doi.org/#2} {#1}\fi
  \endgroup}
\def\mn@eprint#1#2{\mn@eprint@#1:#2::\@nil}
\def\mn@eprint@arXiv#1{\href {http://arxiv.org/abs/#1} {{\tt arXiv:#1}}}
\def\mn@eprint@dblp#1{\href {http://dblp.uni-trier.de/rec/bibtex/#1.xml}
  {dblp:#1}}
\def\mn@eprint@#1:#2:#3:#4\@nil{\def\@tempa {#1}\def\@tempb {#2}\def\@tempc
  {#3}\ifx \@tempc \@empty \let \@tempc \@tempb \let \@tempb \@tempa \fi \ifx
  \@tempb \@empty \def\@tempb {arXiv}\fi \@ifundefined
  {mn@eprint@\@tempb}{\@tempb:\@tempc}{\expandafter \expandafter \csname
  mn@eprint@\@tempb\endcsname \expandafter{\@tempc}}}

\bibitem[\protect\citeauthoryear{{Alves}, {Lombardi}  \& {Lada}}{{Alves}
  et~al.}{2007}]{Alves_CMF_IMF_obs}
{Alves} J.,  {Lombardi} M.,   {Lada} C.~J.,  2007, \mn@doi [\aap]
  {10.1051/0004-6361:20066389}, \href
  {http://adsabs.harvard.edu/abs/2007A%26A...462L..17A} {462, L17}

\bibitem[\protect\citeauthoryear{{Audit} \& {Hennebelle}}{{Audit} \&
  {Hennebelle}}{2010}]{audit:2010.gmc.massfunctions}
{Audit} E.,  {Hennebelle} P.,  2010, \mn@doi [\aap]
  {10.1051/0004-6361/200912695}, \href
  {http://adsabs.harvard.edu/abs/2010A%26A...511A..76A} {511, A76+}

\bibitem[\protect\citeauthoryear{{Ballesteros-Paredes}, {Vazquez-Semadeni},
  {Gazol}, {Hartmann}, {Heitsch}  \& {Colin}}{{Ballesteros-Paredes}
  et~al.}{2011}]{ballesteros-paredes:2011.dens.pdf.vs.selfgrav}
{Ballesteros-Paredes} J.,  {Vazquez-Semadeni} E.,  {Gazol} A.,  {Hartmann}
  L.~W.,  {Heitsch} F.,   {Colin} P.,  2011, \mn@doi [\mnras]
  {10.1111/j.1365-2966.2011.19141.x}, \href
  {http://adsabs.harvard.edu/abs/2011arXiv1105.5411B} {416, 1436}

\bibitem[\protect\citeauthoryear{{Bastian}, {Covey}  \& {Meyer}}{{Bastian}
  et~al.}{2010}]{IMF_review}
{Bastian} N.,  {Covey} K.~R.,   {Meyer} M.~R.,  2010, \mn@doi [\araa]
  {10.1146/annurev-astro-082708-101642}, \href
  {http://adsabs.harvard.edu/abs/2010ARA%26A..48..339B} {48, 339}

\bibitem[\protect\citeauthoryear{{Baugh}}{{Baugh}}{1996}]{Baugh_1996_DM_correlation}
{Baugh} C.~M.,  1996, \mn@doi [\mnras] {10.1093/mnras/280.1.267}, \href
  {http://adsabs.harvard.edu/abs/1996MNRAS.280..267B} {280, 267}

\bibitem[\protect\citeauthoryear{{Bik}, {Lamers}, {Bastian}, {Panagia}  \&
  {Romaniello}}{{Bik} et~al.}{2003}]{Bik_2003_cluster_IMF}
{Bik} A.,  {Lamers} H.~J.~G.~L.~M.,  {Bastian} N.,  {Panagia} N.,
  {Romaniello} M.,  2003, \mn@doi [\aap] {10.1051/0004-6361:20021384}, \href
  {http://adsabs.harvard.edu/abs/2003A%26A...397..473B} {397, 473}

\bibitem[\protect\citeauthoryear{{Bolatto}, {Leroy}, {Rosolowsky}, {Walter}  \&
  {Blitz}}{{Bolatto} et~al.}{2008}]{Bolatto_2008}
{Bolatto} A.~D.,  {Leroy} A.~K.,  {Rosolowsky} E.,  {Walter} F.,   {Blitz} L.,
  2008, \mn@doi [\apj] {10.1086/591513}, \href
  {http://adsabs.harvard.edu/abs/2008ApJ...686..948B} {686, 948}

\bibitem[\protect\citeauthoryear{{Bond}, {Cole}, {Efstathiou}  \&
  {Kaiser}}{{Bond} et~al.}{1991}]{Bond_extended_PS}
{Bond} J.~R.,  {Cole} S.,  {Efstathiou} G.,   {Kaiser} N.,  1991, \mn@doi
  [\apj] {10.1086/170520}, \href
  {http://adsabs.harvard.edu/abs/1991ApJ...379..440B} {379, 440}

\bibitem[\protect\citeauthoryear{{Bonnell}, {Larson}  \& {Zinnecker}}{{Bonnell}
  et~al.}{2007}]{bonnell_2007_competitive_accretion_imf}
{Bonnell} I.~A.,  {Larson} R.~B.,   {Zinnecker} H.,  2007, Protostars and
  Planets V, \href {http://adsabs.harvard.edu/abs/2007prpl.conf..149B} {pp
  149--164}

\bibitem[\protect\citeauthoryear{{Burkhart}, {Collins}  \&
  {Lazarian}}{{Burkhart} et~al.}{2015}]{Burkhart_cloud_sim_2015}
{Burkhart} B.,  {Collins} D.~C.,   {Lazarian} A.,  2015, \mn@doi [\apj]
  {10.1088/0004-637X/808/1/48}, \href
  {http://adsabs.harvard.edu/abs/2015ApJ...808...48B} {808, 48}

\bibitem[\protect\citeauthoryear{{Chappell} \& {Scalo}}{{Chappell} \&
  {Scalo}}{2001}]{Chappell_Scalo_2001_multifractal_ISM}
{Chappell} D.,  {Scalo} J.,  2001, \mn@doi [\apj] {10.1086/320242}, \href
  {http://adsabs.harvard.edu/abs/2001ApJ...551..712C} {551, 712}

\bibitem[\protect\citeauthoryear{{Collins}, {Kritsuk}, {Padoan}, {Li}, {Xu},
  {Ustyugov}  \& {Norman}}{{Collins} et~al.}{2012}]{Collins_2012}
{Collins} D.~C.,  {Kritsuk} A.~G.,  {Padoan} P.,  {Li} H.,  {Xu} H.,
  {Ustyugov} S.~D.,   {Norman} M.~L.,  2012, \mn@doi [\apj]
  {10.1088/0004-637X/750/1/13}, \href
  {http://adsabs.harvard.edu/abs/2012ApJ...750...13C} {750, 13}

\bibitem[\protect\citeauthoryear{{Colombo} et~al.,}{{Colombo}
  et~al.}{2014}]{Colombo_2014_GMC_survey}
{Colombo} D.,  et~al., 2014, \mn@doi [\apj] {10.1088/0004-637X/784/1/3}, \href
  {http://adsabs.harvard.edu/abs/2014ApJ...784....3C} {784, 3}

\bibitem[\protect\citeauthoryear{{Crutcher}}{{Crutcher}}{2012}]{Crutcher_2009_MC_magnetic_fields}
{Crutcher} R.~M.,  2012, \mn@doi [\araa] {10.1146/annurev-astro-081811-125514},
  \href {http://adsabs.harvard.edu/abs/2012ARA%26A..50...29C} {50, 29}

\bibitem[\protect\citeauthoryear{{Elmegreen}}{{Elmegreen}}{1997}]{Elmegreen_1997_IMF_fractal_model}
{Elmegreen} B.~G.,  1997, \mn@doi [\apj] {10.1086/304562}, \href
  {http://adsabs.harvard.edu/abs/1997ApJ...486..944E} {486, 944}

\bibitem[\protect\citeauthoryear{{Elmegreen}}{{Elmegreen}}{2002}]{Elmegreen_2002_cloud_MF_fractal}
{Elmegreen} B.~G.,  2002, \mn@doi [\apj] {10.1086/324384}, \href
  {http://adsabs.harvard.edu/abs/2002ApJ...564..773E} {564, 773}

\bibitem[\protect\citeauthoryear{{Elmegreen} \& {Falgarone}}{{Elmegreen} \&
  {Falgarone}}{1996}]{Elmegreen_1996_fractal}
{Elmegreen} B.~G.,  {Falgarone} E.,  1996, \mn@doi [\apj] {10.1086/178009},
  \href {http://adsabs.harvard.edu/abs/1996ApJ...471..816E} {471, 816}

\bibitem[\protect\citeauthoryear{{Elson}, {Fall}  \& {Freeman}}{{Elson}
  et~al.}{1987}]{Elson1987}
{Elson} R.~A.~W.,  {Fall} S.~M.,   {Freeman} K.~C.,  1987, \mn@doi [\apj]
  {10.1086/165807}, \href {http://adsabs.harvard.edu/abs/1987ApJ...323...54E}
  {323, 54}

\bibitem[\protect\citeauthoryear{{Fall} \& {Chandar}}{{Fall} \&
  {Chandar}}{2012}]{Fall_2012_similarities}
{Fall} S.~M.,  {Chandar} R.,  2012, \mn@doi [\apj]
  {10.1088/0004-637X/752/2/96}, \href
  {http://adsabs.harvard.edu/abs/2012ApJ...752...96F} {752, 96}

\bibitem[\protect\citeauthoryear{{Federrath} \& {Banerjee}}{{Federrath} \&
  {Banerjee}}{2015}]{federrath.2015:density.pdf.and.sfr.in.polytropic.turbulence}
{Federrath} C.,  {Banerjee} S.,  2015, \mn@doi [\mnras] {10.1093/mnras/stv180},
  \href {http://adsabs.harvard.edu/abs/2015MNRAS.448.3297F} {448, 3297}

\bibitem[\protect\citeauthoryear{{Federrath} \& {Klessen}}{{Federrath} \&
  {Klessen}}{2012}]{Federrath_sim_2012}
{Federrath} C.,  {Klessen} R.~S.,  2012, \mn@doi [\apj]
  {10.1088/0004-637X/761/2/156}, \href
  {http://adsabs.harvard.edu/abs/2012ApJ...761..156F} {761, 156}

\bibitem[\protect\citeauthoryear{{Federrath}, {Roman-Duval}, {Klessen},
  {Schmidt}  \& {Mac Low}}{{Federrath}
  et~al.}{2010}]{federrath:2010.obs.vs.sim.turb.compare}
{Federrath} C.,  {Roman-Duval} J.,  {Klessen} R.~S.,  {Schmidt} W.,   {Mac Low}
  M.-M.,  2010, \mn@doi [\aap] {10.1051/0004-6361/200912437}, \href
  {http://adsabs.harvard.edu/abs/2010A%26A...512A..81F} {512, A81+}

\bibitem[\protect\citeauthoryear{{Girichidis}, {Konstandin}, {Whitworth}  \&
  {Klessen}}{{Girichidis} et~al.}{2014}]{Girichidis_density_PDF_2014}
{Girichidis} P.,  {Konstandin} L.,  {Whitworth} A.~P.,   {Klessen} R.~S.,
  2014, \mn@doi [\apj] {10.1088/0004-637X/781/2/91}, \href
  {http://adsabs.harvard.edu/abs/2014ApJ...781...91G} {781, 91}

\bibitem[\protect\citeauthoryear{{Gouliermis}, {Hony}  \&
  {Klessen}}{{Gouliermis} et~al.}{2014}]{Gouliermis_2014}
{Gouliermis} D.~A.,  {Hony} S.,   {Klessen} R.~S.,  2014, \mn@doi [\mnras]
  {10.1093/mnras/stu228}, \href
  {http://adsabs.harvard.edu/abs/2014MNRAS.439.3775G} {439, 3775}

\bibitem[\protect\citeauthoryear{{Grasha} et~al.,}{{Grasha}
  et~al.}{2017}]{grasha_2017_cluster_correlation}
{Grasha} K.,  et~al., 2017, preprint, \href
  {http://adsabs.harvard.edu/abs/2017arXiv170406321G} {} (\mn@eprint {arXiv}
  {1704.06321})

\bibitem[\protect\citeauthoryear{{Grudi{\'c}}, {Hopkins},
  {Faucher-Gigu{\`e}re}, {Quataert}, {Murray}  \& {Kere{\v s}}}{{Grudi{\'c}}
  et~al.}{2016}]{Grudic_2016}
{Grudi{\'c}} M.~Y.,  {Hopkins} P.~F.,  {Faucher-Gigu{\`e}re} C.-A.,  {Quataert}
  E.,  {Murray} N.,   {Kere{\v s}} D.,  2016, preprint, \href
  {http://adsabs.harvard.edu/abs/2016arXiv161205635G} {} (\mn@eprint {arXiv}
  {1612.05635})

\bibitem[\protect\citeauthoryear{{Grudi{\'c}}, {Guszejnov}, {Hopkins},
  {Lamberts}, {Boylan-Kolchin}, {Murray}  \& {Schmitz}}{{Grudi{\'c}}
  et~al.}{2017}]{grudic_2017}
{Grudi{\'c}} M.~Y.,  {Guszejnov} D.,  {Hopkins} P.~F.,  {Lamberts} A.,
  {Boylan-Kolchin} M.,  {Murray} N.,   {Schmitz} D.,  2017, preprint, \href
  {http://adsabs.harvard.edu/abs/2017arXiv170809065G} {} (\mn@eprint {arXiv}
  {1708.09065})

\bibitem[\protect\citeauthoryear{{Guszejnov} \& {Hopkins}}{{Guszejnov} \&
  {Hopkins}}{2016}]{guszejnov_GMC_IMF}
{Guszejnov} D.,  {Hopkins} P.~F.,  2016, \mn@doi [\mnras]
  {10.1093/mnras/stw619}, \href
  {http://adsabs.harvard.edu/abs/2016MNRAS.459....9G} {459, 9}

\bibitem[\protect\citeauthoryear{{Guszejnov}, {Hopkins}  \&
  {Krumholz}}{{Guszejnov} et~al.}{2016a}]{guszejnov_correlation}
{Guszejnov} D.,  {Hopkins} P.~F.,   {Krumholz} M.~R.,  2016a, preprint, \href
  {http://adsabs.harvard.edu/abs/2016arXiv161000772G} {} (\mn@eprint {arXiv}
  {1610.00772})

\bibitem[\protect\citeauthoryear{{Guszejnov}, {Krumholz}  \&
  {Hopkins}}{{Guszejnov} et~al.}{2016b}]{guszejnov_feedback_necessity}
{Guszejnov} D.,  {Krumholz} M.~R.,   {Hopkins} P.~F.,  2016b, \mn@doi [\mnras]
  {10.1093/mnras/stw315}, \href
  {http://adsabs.harvard.edu/abs/2016MNRAS.458..673G} {458, 673}

\bibitem[\protect\citeauthoryear{{Hartmann}}{{Hartmann}}{2002}]{Hartmann2002}
{Hartmann} L.,  2002, \mn@doi [\apj] {10.1086/342657}, \href
  {http://adsabs.harvard.edu/abs/2002ApJ...578..914H} {578, 914}

\bibitem[\protect\citeauthoryear{{Hennebelle} \& {Chabrier}}{{Hennebelle} \&
  {Chabrier}}{2008}]{HC08}
{Hennebelle} P.,  {Chabrier} G.,  2008, \mn@doi [\apj] {10.1086/589916}, \href
  {http://adsabs.harvard.edu/abs/2008ApJ...684..395H} {684, 395}

\bibitem[\protect\citeauthoryear{{Hennebelle} \& {Chabrier}}{{Hennebelle} \&
  {Chabrier}}{2009}]{HC2009}
{Hennebelle} P.,  {Chabrier} G.,  2009, \mn@doi [\apj]
  {10.1088/0004-637X/702/2/1428}, \href
  {http://adsabs.harvard.edu/abs/2009ApJ...702.1428H} {702, 1428}

\bibitem[\protect\citeauthoryear{{Hennebelle} \& {Chabrier}}{{Hennebelle} \&
  {Chabrier}}{2013}]{HC_2013}
{Hennebelle} P.,  {Chabrier} G.,  2013, \mn@doi [\apj]
  {10.1088/0004-637X/770/2/150}, \href
  {http://adsabs.harvard.edu/abs/2013ApJ...770..150H} {770, 150}

\bibitem[\protect\citeauthoryear{{Hennekemper}, {Gouliermis}, {Henning},
  {Brandner}  \& {Dolphin}}{{Hennekemper} et~al.}{2008}]{Hennekemper2008}
{Hennekemper} E.,  {Gouliermis} D.~A.,  {Henning} T.,  {Brandner} W.,
  {Dolphin} A.~E.,  2008, \mn@doi [\apj] {10.1086/524105}, \href
  {http://adsabs.harvard.edu/abs/2008ApJ...672..914H} {672, 914}

\bibitem[\protect\citeauthoryear{{Hopkins}}{{Hopkins}}{2012a}]{excursion_set_ism}
{Hopkins} P.~F.,  2012a, \mn@doi [\mnras] {10.1111/j.1365-2966.2012.20730.x},
  \href {http://adsabs.harvard.edu/abs/2012MNRAS.423.2016H} {423, 2016}

\bibitem[\protect\citeauthoryear{{Hopkins}}{{Hopkins}}{2012b}]{core_IMF}
{Hopkins} P.~F.,  2012b, \mn@doi [\mnras] {10.1111/j.1365-2966.2012.20731.x},
  \href {http://adsabs.harvard.edu/abs/2012MNRAS.423.2037H} {423, 2037}

\bibitem[\protect\citeauthoryear{{Hopkins}}{{Hopkins}}{2013a}]{Hopkins_clustering}
{Hopkins} P.~F.,  2013a, \mn@doi [\mnras] {10.1093/mnras/sts147}, \href
  {http://adsabs.harvard.edu/abs/2013MNRAS.428.1950H} {428, 1950}

\bibitem[\protect\citeauthoryear{{Hopkins}}{{Hopkins}}{2013b}]{general_turbulent_fragment}
{Hopkins} P.~F.,  2013b, \mn@doi [\mnras] {10.1093/mnras/sts704}, \href
  {http://adsabs.harvard.edu/abs/2013MNRAS.430.1653H} {430, 1653}

\bibitem[\protect\citeauthoryear{{Hopkins} et~al.,}{{Hopkins}
  et~al.}{2017}]{Hopkins_fire2}
{Hopkins} P.~F.,  et~al., 2017, preprint, \href
  {http://adsabs.harvard.edu/abs/2017arXiv170206148H} {} (\mn@eprint {arXiv}
  {1702.06148})

\bibitem[\protect\citeauthoryear{{Johnstone} \& {Bally}}{{Johnstone} \&
  {Bally}}{2006}]{Johnstone_Bally_2006_clump_MF}
{Johnstone} D.,  {Bally} J.,  2006, \mn@doi [\apj] {10.1086/508852}, \href
  {http://adsabs.harvard.edu/abs/2006ApJ...653..383J} {653, 383}

\bibitem[\protect\citeauthoryear{{Kainulainen}, {Beuther}, {Henning}  \&
  {Plume}}{{Kainulainen} et~al.}{2009}]{Kainulainen_power_law_tail}
{Kainulainen} J.,  {Beuther} H.,  {Henning} T.,   {Plume} R.,  2009, \mn@doi
  [\aap] {10.1051/0004-6361/200913605}, \href
  {http://adsabs.harvard.edu/abs/2009A%26A...508L..35K} {508, L35}

\bibitem[\protect\citeauthoryear{{Kauffmann}, {White}  \&
  {Guiderdoni}}{{Kauffmann} et~al.}{1993}]{Kauffmann_1993_DM_galaxy_meger}
{Kauffmann} G.,  {White} S.~D.~M.,   {Guiderdoni} B.,  1993, \mn@doi [\mnras]
  {10.1093/mnras/264.1.201}, \href
  {http://adsabs.harvard.edu/abs/1993MNRAS.264..201K} {264, 201}

\bibitem[\protect\citeauthoryear{{Kauffmann}, {Colberg}, {Diaferio}  \&
  {White}}{{Kauffmann} et~al.}{1999}]{Kauffmann_1999_DM_corr_sim}
{Kauffmann} G.,  {Colberg} J.~M.,  {Diaferio} A.,   {White} S.~D.~M.,  1999,
  \mn@doi [\mnras] {10.1046/j.1365-8711.1999.02202.x}, \href
  {http://adsabs.harvard.edu/abs/1999MNRAS.303..188K} {303, 188}

\bibitem[\protect\citeauthoryear{{Klessen} \& {Burkert}}{{Klessen} \&
  {Burkert}}{2001}]{klessen:2001.sf.cloud.pdf}
{Klessen} R.~S.,  {Burkert} A.,  2001, \mn@doi [\apj] {10.1086/319053}, \href
  {http://adsabs.harvard.edu/abs/2001ApJ...549..386K} {549, 386}

\bibitem[\protect\citeauthoryear{{Kramer}, {Stutzki}, {Rohrig}  \&
  {Corneliussen}}{{Kramer} et~al.}{1998}]{Kramer_1998_clumps_MF}
{Kramer} C.,  {Stutzki} J.,  {Rohrig} R.,   {Corneliussen} U.,  1998, \aap,
  \href {http://adsabs.harvard.edu/abs/1998A%26A...329..249K} {329, 249}

\bibitem[\protect\citeauthoryear{{Kraus} \& {Hillenbrand}}{{Kraus} \&
  {Hillenbrand}}{2008}]{Kraus2008}
{Kraus} A.~L.,  {Hillenbrand} L.~A.,  2008, \mn@doi [\apjl] {10.1086/593012},
  \href {http://adsabs.harvard.edu/abs/2008ApJ...686L.111K} {686, L111}

\bibitem[\protect\citeauthoryear{{Kritsuk}, {Norman}  \& {Padoan}}{{Kritsuk}
  et~al.}{2006}]{Kristuk_2006_turbulence}
{Kritsuk} A.~G.,  {Norman} M.~L.,   {Padoan} P.,  2006, \mn@doi [\apjl]
  {10.1086/500688}, \href {http://adsabs.harvard.edu/abs/2006ApJ...638L..25K}
  {638, L25}

\bibitem[\protect\citeauthoryear{{Kritsuk}, {Norman}  \& {Wagner}}{{Kritsuk}
  et~al.}{2011}]{Kritsuk_dens_lognormal}
{Kritsuk} A.~G.,  {Norman} M.~L.,   {Wagner} R.,  2011, \mn@doi [\apjl]
  {10.1088/2041-8205/727/1/L20}, \href
  {http://adsabs.harvard.edu/abs/2011ApJ...727L..20K} {727, L20}

\bibitem[\protect\citeauthoryear{{Kritsuk}, {Lee}  \& {Norman}}{{Kritsuk}
  et~al.}{2013}]{Kritsuk_larson_supersonic_origin}
{Kritsuk} A.~G.,  {Lee} C.~T.,   {Norman} M.~L.,  2013, \mn@doi [\mnras]
  {10.1093/mnras/stt1805}, \href
  {http://adsabs.harvard.edu/abs/2013MNRAS.436.3247K} {436, 3247}

\bibitem[\protect\citeauthoryear{{Krumholz}}{{Krumholz}}{2014}]{SF_big_problems}
{Krumholz} M.~R.,  2014, \mn@doi [\physrep] {10.1016/j.physrep.2014.02.001},
  \href {http://adsabs.harvard.edu/abs/2014PhR...539...49K} {539, 49}

\bibitem[\protect\citeauthoryear{{Lacey} \& {Cole}}{{Lacey} \&
  {Cole}}{1993}]{Lacey_Cole_1993_DM_merger}
{Lacey} C.,  {Cole} S.,  1993, \mn@doi [\mnras] {10.1093/mnras/262.3.627},
  \href {http://adsabs.harvard.edu/abs/1993MNRAS.262..627L} {262, 627}

\bibitem[\protect\citeauthoryear{{Larsen}}{{Larsen}}{2004}]{larsen_2004}
{Larsen} S.~S.,  2004, \mn@doi [\aap] {10.1051/0004-6361:20034533}, \href
  {http://adsabs.harvard.edu/abs/2004A%26A...416..537L} {416, 537}

\bibitem[\protect\citeauthoryear{{Larson}}{{Larson}}{1969}]{Larson_1969}
{Larson} R.~B.,  1969, \mn@doi [\mnras] {10.1093/mnras/145.3.271}, \href
  {http://adsabs.harvard.edu/abs/1969MNRAS.145..271L} {145, 271}

\bibitem[\protect\citeauthoryear{{Larson}}{{Larson}}{1981}]{Larson_law}
{Larson} R.~B.,  1981, \mnras, \href
  {http://adsabs.harvard.edu/abs/1981MNRAS.194..809L} {194, 809}

\bibitem[\protect\citeauthoryear{{Larson}}{{Larson}}{1982}]{Larson_1982}
{Larson} R.~B.,  1982, \mn@doi [\mnras] {10.1093/mnras/200.2.159}, \href
  {http://adsabs.harvard.edu/abs/1982MNRAS.200..159L} {200, 159}

\bibitem[\protect\citeauthoryear{{Lin} et~al.,}{{Lin}
  et~al.}{2017}]{Lin_cloud_structure_2017}
{Lin} Y.,  et~al., 2017, \mn@doi [\apj] {10.3847/1538-4357/aa6c67}, \href
  {http://adsabs.harvard.edu/abs/2017ApJ...840...22L} {840, 22}

\bibitem[\protect\citeauthoryear{{Lombardi}, {Bouy}, {Alves}  \&
  {Lada}}{{Lombardi} et~al.}{2014}]{Lombardi_2014_column_density}
{Lombardi} M.,  {Bouy} H.,  {Alves} J.,   {Lada} C.~J.,  2014, \mn@doi [\aap]
  {10.1051/0004-6361/201323293}, \href
  {http://adsabs.harvard.edu/abs/2014A%26A...566A..45L} {566, A45}

\bibitem[\protect\citeauthoryear{{Lynden-Bell}}{{Lynden-Bell}}{1967}]{Lynden_Bell_1967}
{Lynden-Bell} D.,  1967, \mn@doi [\mnras] {10.1093/mnras/136.1.101}, \href
  {http://adsabs.harvard.edu/abs/1967MNRAS.136..101L} {136, 101}

\bibitem[\protect\citeauthoryear{{Mackey} \& {Gilmore}}{{Mackey} \&
  {Gilmore}}{2003a}]{Mackey_Gilmore_2003}
{Mackey} A.~D.,  {Gilmore} G.~F.,  2003a, \mn@doi [\mnras]
  {10.1046/j.1365-8711.2003.06021.x}, \href
  {http://adsabs.harvard.edu/abs/2003MNRAS.338...85M} {338, 85}

\bibitem[\protect\citeauthoryear{{Mackey} \& {Gilmore}}{{Mackey} \&
  {Gilmore}}{2003b}]{Mackey_Gilmore_2003_SMC}
{Mackey} A.~D.,  {Gilmore} G.~F.,  2003b, \mn@doi [\mnras]
  {10.1046/j.1365-8711.2003.06022.x}, \href
  {http://adsabs.harvard.edu/abs/2003MNRAS.338..120M} {338, 120}

\bibitem[\protect\citeauthoryear{{Massey}}{{Massey}}{2003}]{Massey_2003_IMF_slope}
{Massey} P.,  2003, \mn@doi [\araa] {10.1146/annurev.astro.41.071601.170033},
  \href {http://adsabs.harvard.edu/abs/2003ARA%26A..41...15M} {41, 15}

\bibitem[\protect\citeauthoryear{{Murray}}{{Murray}}{1973}]{Murray_supersonic_Burgers_analytic}
{Murray} J.~D.,  1973, \mn@doi [Journal of Fluid Mechanics]
  {10.1017/S0022112073001564}, \href
  {http://adsabs.harvard.edu/abs/1973JFM....59..263M} {59, 263}

\bibitem[\protect\citeauthoryear{{Nakajima}, {Tachihara}, {Hanawa}  \&
  {Nakano}}{{Nakajima} et~al.}{1998}]{Nakajima1998}
{Nakajima} Y.,  {Tachihara} K.,  {Hanawa} T.,   {Nakano} M.,  1998, \mn@doi
  [\apj] {10.1086/305493}, \href
  {http://adsabs.harvard.edu/abs/1998ApJ...497..721N} {497, 721}

\bibitem[\protect\citeauthoryear{{Navarro}, {Frenk}  \& {White}}{{Navarro}
  et~al.}{1996}]{NFW_1996}
{Navarro} J.~F.,  {Frenk} C.~S.,   {White} S.~D.~M.,  1996, \mn@doi [\apj]
  {10.1086/177173}, \href {http://adsabs.harvard.edu/abs/1996ApJ...462..563N}
  {462, 563}

\bibitem[\protect\citeauthoryear{{Offner}, {Clark}, {Hennebelle}, {Bastian},
  {Bate}, {Hopkins}, {Moraux}  \& {Whitworth}}{{Offner}
  et~al.}{2014}]{IMF_universality}
{Offner} S.~S.~R.,  {Clark} P.~C.,  {Hennebelle} P.,  {Bastian} N.,  {Bate}
  M.~R.,  {Hopkins} P.~F.,  {Moraux} E.,   {Whitworth} A.~P.,  2014, \mn@doi
  [Protostars and Planets VI] {10.2458/azu_uapress_9780816531240-ch003}, \href
  {http://adsabs.harvard.edu/abs/2014prpl.conf...53O} {pp 53--75}

\bibitem[\protect\citeauthoryear{{Ostriker}, {Gammie}  \& {Stone}}{{Ostriker}
  et~al.}{1999}]{ostriker:1999.density.pdf}
{Ostriker} E.~C.,  {Gammie} C.~F.,   {Stone} J.~M.,  1999, \mn@doi [\apj]
  {10.1086/306842}, \href {http://adsabs.harvard.edu/abs/1999ApJ...513..259O}
  {513, 259}

\bibitem[\protect\citeauthoryear{{Padoan} \& {Nordlund}}{{Padoan} \&
  {Nordlund}}{2002}]{Padoan_Nordlund_2002_IMF}
{Padoan} P.,  {Nordlund} {\AA}.,  2002, \mn@doi [\apj] {10.1086/341790}, \href
  {http://adsabs.harvard.edu/abs/2002ApJ...576..870P} {576, 870}

\bibitem[\protect\citeauthoryear{{Padoan}, {Nordlund}  \& {Jones}}{{Padoan}
  et~al.}{1997}]{Padoan_theory}
{Padoan} P.,  {Nordlund} A.,   {Jones} B.~J.~T.,  1997, \mnras, \href
  {http://adsabs.harvard.edu/abs/1997MNRAS.288..145P} {288, 145}

\bibitem[\protect\citeauthoryear{{Penston}}{{Penston}}{1969}]{Penston_1969}
{Penston} M.~V.,  1969, \mn@doi [\mnras] {10.1093/mnras/144.4.425}, \href
  {http://adsabs.harvard.edu/abs/1969MNRAS.144..425P} {144, 425}

\bibitem[\protect\citeauthoryear{{Portegies Zwart}, {McMillan}  \&
  {Gieles}}{{Portegies Zwart} et~al.}{2010}]{Portegies_young_clusters}
{Portegies Zwart} S.~F.,  {McMillan} S.~L.~W.,   {Gieles} M.,  2010, \mn@doi
  [\araa] {10.1146/annurev-astro-081309-130834}, \href
  {http://adsabs.harvard.edu/abs/2010ARA%26A..48..431P} {48, 431}

\bibitem[\protect\citeauthoryear{{Press} \& {Schechter}}{{Press} \&
  {Schechter}}{1974}]{PressSchechter}
{Press} W.~H.,  {Schechter} P.,  1974, \mn@doi [\apj] {10.1086/152650}, \href
  {http://adsabs.harvard.edu/abs/1974ApJ...187..425P} {187, 425}

\bibitem[\protect\citeauthoryear{{Rosolowsky}}{{Rosolowsky}}{2005}]{Rosolowsky_2005_GMC}
{Rosolowsky} E.,  2005, \mn@doi [\pasp] {10.1086/497582}, \href
  {http://adsabs.harvard.edu/abs/2005PASP..117.1403R} {117, 1403}

\bibitem[\protect\citeauthoryear{{Ryon} et~al.,}{{Ryon}
  et~al.}{2015}]{ryon_2015_M83_clusters}
{Ryon} J.~E.,  et~al., 2015, \mn@doi [\mnras] {10.1093/mnras/stv1282}, \href
  {http://adsabs.harvard.edu/abs/2015MNRAS.452..525R} {452, 525}

\bibitem[\protect\citeauthoryear{{Sadavoy} et~al.,}{{Sadavoy}
  et~al.}{2010}]{Sadavoy_observed_CMF}
{Sadavoy} S.~I.,  et~al., 2010, \mn@doi [\apj] {10.1088/0004-637X/710/2/1247},
  \href {http://adsabs.harvard.edu/abs/2010ApJ...710.1247S} {710, 1247}

\bibitem[\protect\citeauthoryear{{Salpeter}}{{Salpeter}}{1955}]{Salpeter_slope}
{Salpeter} E.~E.,  1955, \mn@doi [\apj] {10.1086/145971}, \href
  {http://adsabs.harvard.edu/abs/1955ApJ...121..161S} {121, 161}

\bibitem[\protect\citeauthoryear{{Scalo}, {Vazquez-Semadeni}, {Chappell}  \&
  {Passot}}{{Scalo} et~al.}{1998}]{scalo:1998.turb.density.pdf}
{Scalo} J.,  {Vazquez-Semadeni} E.,  {Chappell} D.,   {Passot} T.,  1998,
  \mn@doi [\apj] {10.1086/306099}, \href
  {http://adsabs.harvard.edu/abs/1998ApJ...504..835S} {504, 835}

\bibitem[\protect\citeauthoryear{{Schneider} et~al.,}{{Schneider}
  et~al.}{2013}]{Schneider_2013_orionB_density}
{Schneider} N.,  et~al., 2013, \mn@doi [\apjl] {10.1088/2041-8205/766/2/L17},
  \href {http://adsabs.harvard.edu/abs/2013ApJ...766L..17S} {766, L17}

\bibitem[\protect\citeauthoryear{{Schneider} et~al.,}{{Schneider}
  et~al.}{2015a}]{Schneider_two_power_law_2015}
{Schneider} N.,  et~al., 2015a, \mn@doi [\mnras] {10.1093/mnrasl/slv101}, \href
  {http://adsabs.harvard.edu/abs/2015MNRAS.453L..41S} {453, L41}

\bibitem[\protect\citeauthoryear{{Schneider} et~al.,}{{Schneider}
  et~al.}{2015b}]{Schneider_2015}
{Schneider} N.,  et~al., 2015b, \mn@doi [\aap] {10.1051/0004-6361/201424375},
  \href {http://adsabs.harvard.edu/abs/2015A%26A...578A..29S} {578, A29}

\bibitem[\protect\citeauthoryear{{Scoville}, {Yun}, {Sanders}, {Clemens}  \&
  {Waller}}{{Scoville} et~al.}{1987}]{Scoville_1987_mass_linewidth_size}
{Scoville} N.~Z.,  {Yun} M.~S.,  {Sanders} D.~B.,  {Clemens} D.~P.,   {Waller}
  W.~H.,  1987, \mn@doi [\apjs] {10.1086/191185}, \href
  {http://adsabs.harvard.edu/abs/1987ApJS...63..821S} {63, 821}

\bibitem[\protect\citeauthoryear{{Simon}}{{Simon}}{1997}]{Simon1997}
{Simon} M.,  1997, \mn@doi [\apjl] {10.1086/310678}, \href
  {http://adsabs.harvard.edu/abs/1997ApJ...482L..81S} {482, L81}

\bibitem[\protect\citeauthoryear{{So{\l}tan} \& {Chodorowski}}{{So{\l}tan} \&
  {Chodorowski}}{2015}]{Soltan_2015_galaxy_correlation}
{So{\l}tan} A.~M.,  {Chodorowski} M.~J.,  2015, \mn@doi [\mnras]
  {10.1093/mnras/stv1664}, \href
  {http://adsabs.harvard.edu/abs/2015MNRAS.453.1013S} {453, 1013}

\bibitem[\protect\citeauthoryear{{Squire} \& {Hopkins}}{{Squire} \&
  {Hopkins}}{2017}]{squire.hopkins:turb.density.pdf}
{Squire} J.,  {Hopkins} P.~F.,  2017, \mnras, in press, arXiv:1702.07731, \href
  {http://adsabs.harvard.edu/abs/2017arXiv170207731S} {}

\bibitem[\protect\citeauthoryear{{Stanke}, {Smith}, {Gredel}  \&
  {Khanzadyan}}{{Stanke} et~al.}{2006}]{Stanke2006}
{Stanke} T.,  {Smith} M.~D.,  {Gredel} R.,   {Khanzadyan} T.,  2006, \mn@doi
  [\aap] {10.1051/0004-6361:20041331}, \href
  {http://adsabs.harvard.edu/abs/2006A%26A...447..609S} {447, 609}

\bibitem[\protect\citeauthoryear{{Stutzki}, {Bensch}, {Heithausen}, {Ossenkopf}
   \& {Zielinsky}}{{Stutzki} et~al.}{1998}]{Stutzki_1998_fractal_ISM}
{Stutzki} J.,  {Bensch} F.,  {Heithausen} A.,  {Ossenkopf} V.,   {Zielinsky}
  M.,  1998, \aap, \href {http://adsabs.harvard.edu/abs/1998A%26A...336..697S}
  {336, 697}

\bibitem[\protect\citeauthoryear{{V{\'a}zquez-Semadeni} \&
  {Garc{\'{\i}}a}}{{V{\'a}zquez-Semadeni} \&
  {Garc{\'{\i}}a}}{2001}]{vazquez-semadeni:2001.nh.pdf.gmc}
{V{\'a}zquez-Semadeni} E.,  {Garc{\'{\i}}a} N.,  2001, \mn@doi [\apj]
  {10.1086/321688}, \href {http://adsabs.harvard.edu/abs/2001ApJ...557..727V}
  {557, 727}

\bibitem[\protect\citeauthoryear{{Warren}, {Abazajian}, {Holz}  \&
  {Teodoro}}{{Warren} et~al.}{2006}]{Warren_2006_DM_MF}
{Warren} M.~S.,  {Abazajian} K.,  {Holz} D.~E.,   {Teodoro} L.,  2006, \mn@doi
  [\apj] {10.1086/504962}, \href
  {http://adsabs.harvard.edu/abs/2006ApJ...646..881W} {646, 881}

\bibitem[\protect\citeauthoryear{{Zhang} \& {Fall}}{{Zhang} \&
  {Fall}}{1999}]{Zhang_Falll_1999}
{Zhang} Q.,  {Fall} S.~M.,  1999, \mn@doi [\apjl] {10.1086/312412}, \href
  {http://adsabs.harvard.edu/abs/1999ApJ...527L..81Z} {527, L81}

\bibitem[\protect\citeauthoryear{{Zhang}, {Fall}  \& {Whitmore}}{{Zhang}
  et~al.}{2001}]{Zhang_Fall_2001_cluster_correlation}
{Zhang} Q.,  {Fall} S.~M.,   {Whitmore} B.~C.,  2001, \mn@doi [\apj]
  {10.1086/322278}, \href {http://adsabs.harvard.edu/abs/2001ApJ...561..727Z}
  {561, 727}

\bibitem[\protect\citeauthoryear{{Zinnecker}}{{Zinnecker}}{1982}]{Zinnecker_1982_competitive_accretion}
{Zinnecker} H.,  1982, \mn@doi [Annals of the New York Academy of Sciences]
  {10.1111/j.1749-6632.1982.tb43399.x}, \href
  {http://adsabs.harvard.edu/abs/1982NYASA.395..226Z} {395, 226}

\makeatother
\end{thebibliography}

\appendix

\section{Fractal Dimension and the Correlation Function}\label{sec:fractal}

in this paper we use the fractal $D$ dimension, which we define as
\be
\label{eq:fractal_dim_def}
D\frac{\dderiv\ln N(r)}{\dderiv \ln r},
\ee 
where $N(r)$ is the average number of objects within $r$ distance of a reference object.

In isotropic systems the fractal dimension is related to the $\xi_d(r)$ $d$ dimensional correlation function, for which we use the standard definition of
\begin{eqnarray}
P_d(r,dr)=\frac{N(r,\dderiv r)}{n, \dderiv V_d(r)}\nonumber\\
1+\xi_d(r)=\lim\limits_{\dderiv r \to 0} P_d(r,dr),
\label{eq:correlation}
\end{eqnarray}
where $N(r,\dderiv r)$ is the average number of objects whose $d$ dimensional distance from a reference object is $\in [r,r+\dderiv r]$, $n$ is the density of objects, $V_d(r)$ is the volume of a $d$-sphere so $\dderiv V_d\propto r^{n-1}\dderiv r$. 

Assuming $\xi_d\gg 1$ we get
\be 
\xi_d(r)\propto \lim\limits_{\dderiv r \to 0}\frac{ N(r,\dderiv r)}{r^{d-1}\dderiv r}=r^{1-d}\frac{\dderiv N(r)}{\dderiv r}=\frac{\dderiv N(r)}{\dderiv \ln r}r^{-d}.
\ee
Let us also assume that $N(r)$ is a power-law (this is true in scale-free systems like the ones in this paper), which yields
\be 
\label{eq:corr_d_general}
\xi_d(r)\propto r^{D-d}.
\ee

\end{document}